\documentclass[journal=acsomega,manuscript=article]{achemso}

\usepackage{chemformula} 
\usepackage[T1]{fontenc} 
\usepackage{subcaption}

\author{Mohammed AlAloul}
\email{maa9328@nyu.edu}
\author{Mahmoud Rasras}
\affiliation[New York University Abu Dhabi]
{Photonics Research Lab, Department of Electrical and Computer Engineering, New York University Abu Dhabi, Abu Dhabi, UAE}

\title[An \textsf{achemso} demo]
  {Plasmon-enhanced graphene photodetector with CMOS-compatible titanium nitride}

\abbreviations{IR,NMR,UV}

\begin{document}

\begin{singlespace}


\begin{abstract}

  Graphene has emerged as an ultrafast optoelectronic material for on-chip photodetector applications. The 2D nature of graphene enables its facile integration with complementary metal-oxide semiconductor (CMOS) microelectronics and silicon photonics, yet graphene absorbs only $\sim$2.3\% of light. Plasmonic metals can enhance the responsivity of graphene photodetectors, but may result in CMOS-incompatible devices, depending on the choice of metal. Here, we propose a plasmon-enhanced photothermoelectric graphene detector using CMOS-compatible titanium nitride (TiN) on the silicon-on-insulator (SOI) platform. The device performance is quantified by its responsivity, operation speed, and noise equivalent power. Its bandwidth exceeds 100$\,$GHz, and it exhibits a nearly flat photoresponse across the telecom C-band. The photodetector responsivity is as high as 1.4$\,$A/W (1.1$\,$A/W external) at an ultra-compact length of 3.5$\,$\textmu m, which is the most compact footprint reported for a graphene-based waveguide photodetector. Furthermore, it operates at zero-bias, consumes zero energy, and has an ultra-low intrinsic noise equivalent power (NEP$\,$\textless$\,25\:\text{pW/}\sqrt{\text{Hz}}$).
  
\end{abstract}

\section{Introduction}

The integration of optical interconnects with CMOS microelectronics for ultra-high-speed links has become an industrial necessity \cite{Sousa2016TheFO}. Unlike lossy metallic links, integrated optical interconnects can sustain the transmission of ultrafast signals with losses as low as 0.1 dB/cm using the silicon-on-insulator (SOI) platform \cite{SOI}. Moreover, silicon photonics and CMOS microelectronics are both based on silicon, and can in principle function hand-in-hand within a single chip. However, the size disparity between diffraction-limited photonics and advanced CMOS technology nodes, in addition to fabrication and co-integration challenges impede this progress. Introducing plasmonic metals to photonic devices to excite surface plasmon polaritons (SPPs), is one viable solution to this problem, at least for some functionalities, such as photodetection. SPPs are electromagnetic (EM) excitations that exist at the interface between a metal and a dielectric (or a semiconductor). These unique EM excitations result in a significant enhancement of the EM field at the metal-dielectric interface, where light can also be confined beyond its diffraction limit \cite{SPP}. 

Optical interconnects consist of a transmitter, waveguide, and a receiver. The receiver contains a photodetector that converts optical signals to electrical ones. A photodetector is differentiated by its responsivity, speed, footprint, dark current, energy consumption, and ease of integration with standard industrial platforms. Most photodetectors employed in silicon photonics are based on Germanium \cite{2D}. However, these photodetectors are either resistance-capacitance (RC) product or carrier transit-time limited. Incorporating plasmonic waveguides and applying a large bias voltage can push the speed of Germanium photodetectors up to 110 GHz \cite{MSM}, however high energy consumption and large dark currents are associated with this boost. Plasmonic Schottky photodetectors have been proposed for their ease of integration with silicon photonics, and their ability to absorb photons at telecom wavelengths, with a demonstrated responsivity of 0.37$\,$A/W at 3$\,$V bias \cite{Schottky}. Nonetheless, huge dark currents are associated with plasmonic Schottky photodetectors, which could hinder their implementation in applications where high signal-to-noise (SNR) ratios are essential, e.g. telecom. III-V compound semiconductors, that are heterogeneously integrated on silicon, can exhibit a decent performance in terms of speed and responsivity \cite{InP-review}, yet their performance is degraded when integrated with CMOS microelectronics due to packaging parasitics. Besides that, heterogeneous integration techniques are expensive \cite{Ge-on-SOI, 8550947}. Furthermore, the monolithic integration of III-V compound semiconductors with silicon is hindered by the the large mismatches in their lattice constants and thermal expansion coefficients \cite{seamless}. 

Two-dimensional (2D) materials recently emerged as alternative active materials for modulation and photodetection with a special set of inherent features including high-speed, small footprint, low-cost manufacturing, low-power consumption, and CMOS-compatibility \cite{2D}. More specifically, graphene photodetectors gained extraordinary attention for their ultrafast speed and broadband absorption, despite the innate, relatively low optical absorption of graphene \cite{naturegraphene, graphenereview, gan2013chip, wang2013high, xie2018graphene, di2018graphene, youngblood2014multifunctional, wang2015graphene}. Several techniques were proposed to boost the responsivity of high-speed graphene photodetectors including waveguide-integrated configurations where light continuously interacts with the graphene sheet as it propagates through the waveguide, combining graphene with other 2D materials in composite heterostructures, and enhancing graphene's absorption by plasmonic means. So far, the demonstrated waveguide-integrated plasmonic graphene photodetectors are all based on gold \cite{plasmonic1, plasmonic2, plasmonic3, plasmonic4}, which in spite of its outstanding plasmonic performance, is not CMOS-compatible \cite{gold}. 

In this work, we propose an on-chip, compact footprint, ultrafast and high responsivity plasmon-enhanced graphene photodetector that operates at the telecom C-band, and which can be realized using CMOS-compatible processes. The proposed device employs titanium nitride (TiN) as the plasmonic material. TiN is a refractory metal nitride that has optical properties very similar to gold \cite{reviewB}, an electrical conductivity higher than Titanium \cite{TiNconductivity}, yet unlike gold, TiN is CMOS-compatible \cite{patsalas, embedded, cmosTiN}. The device performance is studied and quantified in terms of its responsivity, operation speed, energy consumption, and noise equivalent power.

\section{Device Structure}

\begin{figure}
\begin{subfigure}{.49\textwidth}
  \centering
  \includegraphics[width=1\linewidth]{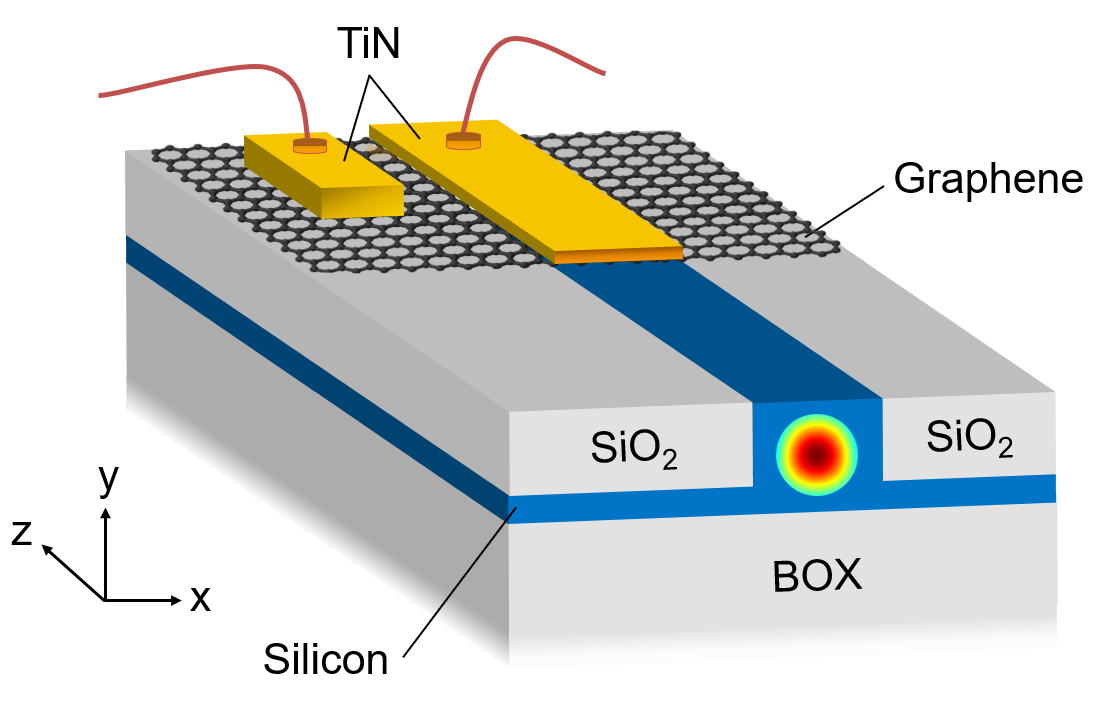}
  \caption{}
  \label{fig:sfig2}
\end{subfigure}%
\begin{subfigure}{.49\textwidth}
  \centering
  \includegraphics[width=1\linewidth]{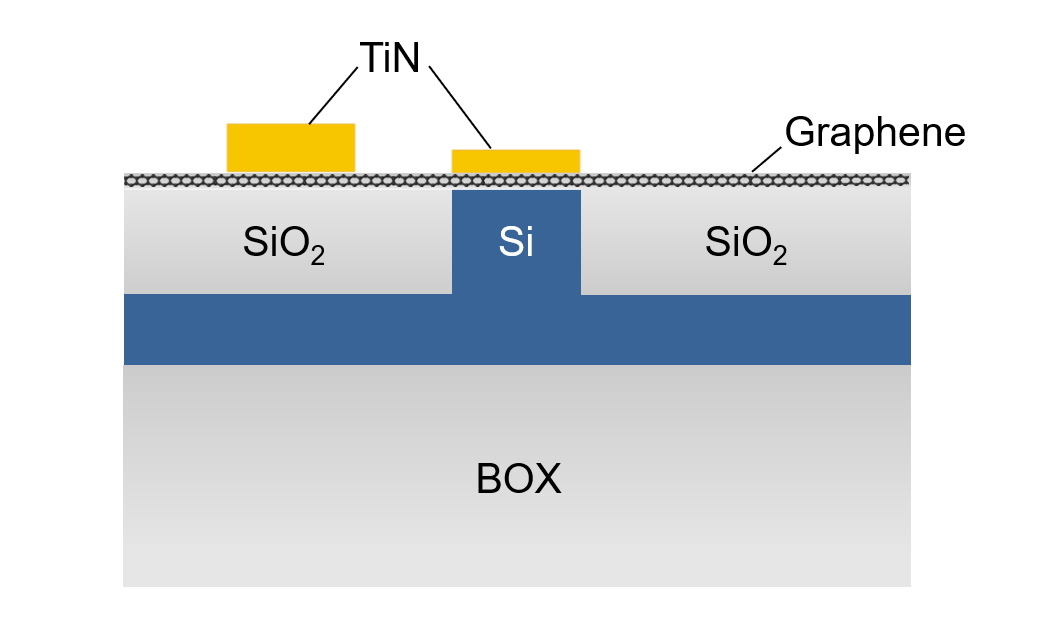}
  \caption{}
  \label{fig:sfig}
\end{subfigure}
\caption{(a) On-chip photodetector structure. (b) Front view of the photodetector with the hexagonal boron nitride (hBN) spacer layer.}
\label{fig:1}
\end{figure}

The structure of the on-chip photodetector is illustrated in Fig \ref{fig:1}. A silicon rib waveguide on top of a 2$\,$\textmu m bottom oxide (BOX) layer guides the incoming light to the photodetector section at the terminating end of the optical link, where the optical signal is absorbed and reverted to its electrical form. The waveguide supports a transverse magnetic (TM)-mode. Within the photodetector section, silicon dioxide (SiO$_{2}$) is deposited on the sides of the waveguide ridge to facilitate the placement of the graphene monolayer on top of the waveguide. The generated electrical signal is collected through the two TiN films that are deposited on top of graphene. Besides collecting the signal, the center TiN film plasmonically enhances the interaction of the propagating TM-mode with graphene, which boosts the optical absorption of the latter, resulting in a high photoresponse at an ultra-compact device length of 3.5$\,$\textmu m. Details of the device geometry, optimization procedure, propagation modes, plasmon-induced losses, and the impact of potential fabrication variations are provided in supplementary section 1.

\section{Operation Principle}

\label{sec:operation}

Graphene detects light mainly through three physical effects: the photovoltaic \cite{gate1, gate2, gate3}, bolometric \cite{bp1, bp2, bp3}, and photo-thermoelectric (PTE) \cite{pte1, pte2, pte3} effects. The bolometric effect is only observed for devices under bias \cite{photoconductivity}, which is not the case for our device. The PTE effect dominates over the photovoltaic effect at zero bias \cite{acshotcarrier}; thus, the photodetection process in our device is predominantly determined by the PTE effect. This effect is based on the phenomenon of photogenerated hot carriers in graphene. Due to the unique conical dispersion of graphene, the density of states fades away at the Dirac point. As a consequence, carriers have a low heat capacity near the Dirac point, and when excited, they immediately scatter with other carriers within a few tens of femtoseconds. These carrier-carrier scattering events result in an ephemeral Fermi-Dirac distribution of hot thermalized carriers, which can be described by a chemical potential (\textmu) and a carrier temperature (T$_{c}$) \cite{carrierScattering}. Afterwards, the hot thermalized carriers cool down in picoseconds by emitting optical and acoustic phonons, coupling with surface optical phonons, and most importantly through disorder-assisted scattering which dominates at room temperature \cite{disorder-assisted, cooling, charged-impurity}.

\begin{figure}
\begin{subfigure}{.49\textwidth}
  \centering
  \includegraphics[width=1\linewidth]{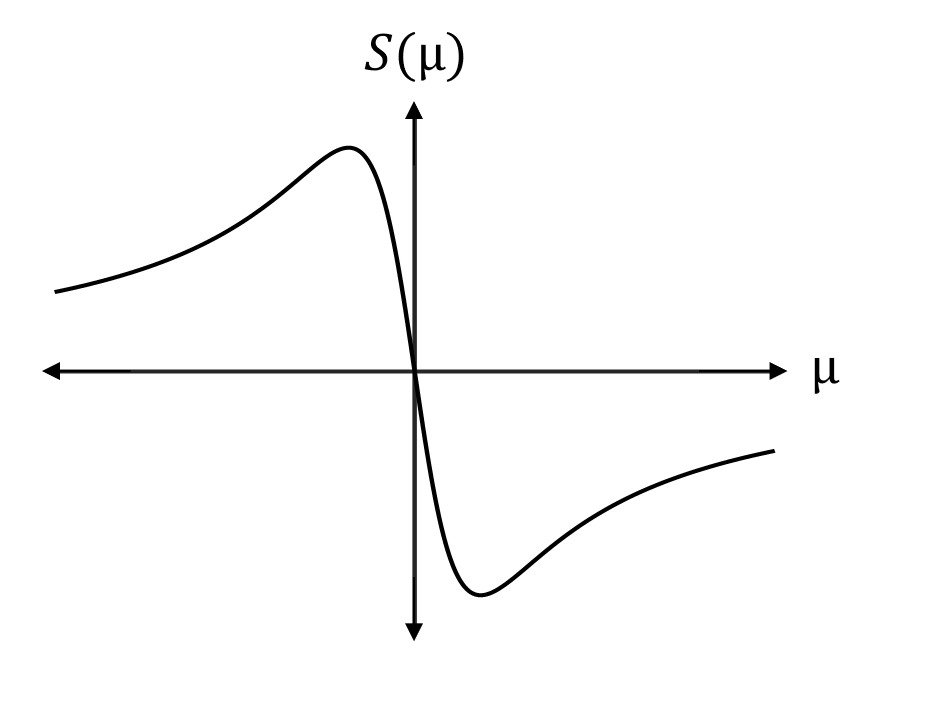}
  \caption{}
  \label{fig:seebeckProp}
\end{subfigure}%
\begin{subfigure}{.49\textwidth}
  \centering
  \includegraphics[width=1\linewidth]{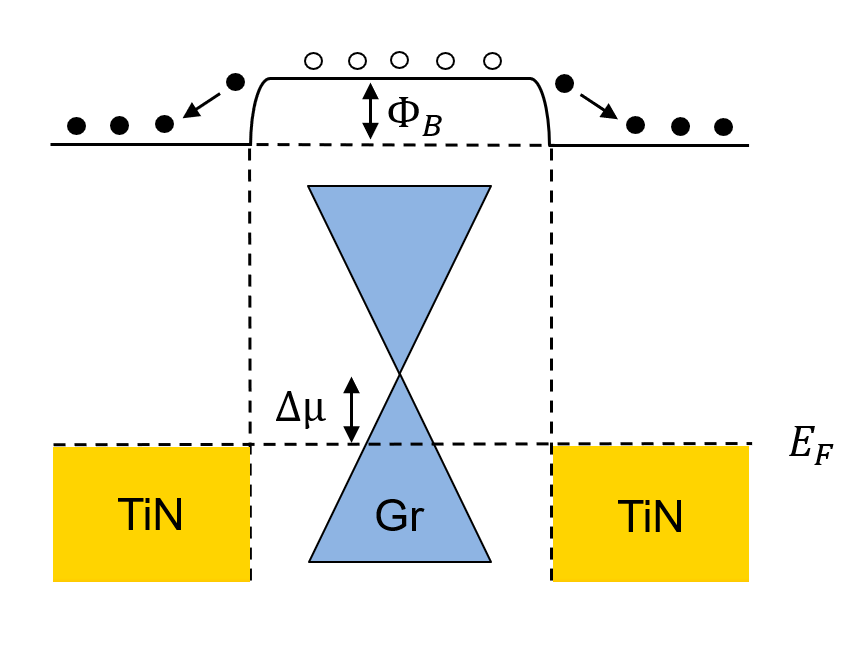}
  \caption{}
  \label{fig:banddiagram}
\end{subfigure}
\caption{(a) Seebeck coefficient as a function of chemical potential. (b) Band diagram representing the doping profile across the graphene sheet; graphene is effectively p-doped by TiN. Black circles and white circles represent electrons and holes, respectively. Gr: graphene.}
\label{fig:random}
\end{figure}

Within the photodetector section, the propagating TM-mode experiences a near-field plasmonic enhancement at the TiN-graphene-Si interface due to the presence of the TiN stripe. This near-field enhancement results in a plasmon-enhanced effective absorption of graphene, and hot carriers are generated as the propagating mode interacts with the graphene sheet. An uneven distribution of hot carriers across the graphene sheet induces a PTE voltage \cite{plasmon-induced, grapheneBook}:

\begin{equation} \label{eq:1}
    V_{PTE} = \int_{0}^{x_{0}} S\: \nabla T_{c} \: dx
\end{equation}

The Seebeck coefficient is a function of the chemical potential, and resembles the curve shown in Fig \ref{fig:seebeckProp}. According to Eq. (\ref{eq:1}), an asymmetric profile of the Seebeck coefficient across the graphene sheet induces a PTE voltage. Graphene is effectively doped when placed in contact with a metal, which in turn affects the Seebeck coefficient and the PTE voltage. The resulting shift in chemical potential ($\Delta$\textmu) is related to the Shottky barrier height ($\Phi_{B})$ at the metal-graphene junction, which is approximated using the Schottky-Mott rule \cite{Pierret1996SemiconductorDF, Monch2017}:

\begin{equation}
    \Phi_{B} = \Phi_{M} - X_{S}
\end{equation}

Where $\Phi_{M}$ is the metal work function, and $X_{S}$ is the electron affinity of the semiconductor given by:

\begin{equation}
    X_{S} = \Phi_{S} - (E_{c} - E_{F})
\end{equation}

Where $\Phi_{S}$ is the work function, $E_{c}$ is the energy of the conduction band edge, and $E_{F}$ is the Fermi energy of the semiconductor. Here we consider TiN, a metal nitride, as the metal, and graphene, a zero-bandgap semiconductor, as the semiconductor. Considering the case of an ideally undoped graphene, gives $(E_{c}-E_{F}) = 0$ and $X_{S} = \Phi_{S}$. Taking work function values of $\sim4.6\:$eV for graphene \cite{WF, NAGHDI20191117}, and 4.2$\:$–$\:4.5\:$eV for TiN \cite{LIMA201286, FILLOT2005248}, gives $\Phi_{B}<0$, leading to electrons flowing from graphene to TiN, or graphene being effectively p-doped by TiN, and the Seebeck coefficient varies as a result (see Fig \ref{fig:banddiagram}). For this device however, the Seebeck coefficient does not contribute to the generated photovoltage, since TiN is placed on both sides of the graphene sheet, giving a symmetric doping profile across it. A difference in the Seebeck coefficient across the graphene sheet is required to generate a $V_{PTE}$ for a fixed $T_{c}$ profile based on \cite{acshotcarrier}:

\begin{equation}
    V_{PTE} = (S_{1} - S_{2}) \: \Delta T \, , \,\,\,\,\, \Delta T = T_{c} - T_{0}
\end{equation}

Where $S_{1}$ and $S_{2}$ represent the Seebeck coefficient at each side of the graphene sheet, and $T_{0}$ is the lattice temperature. Therefore, the photodetector operates solely based on the plasmonic thermoelectric effect. The previous analysis assumes an ideal metal-semiconductor interface. In practice, $\Phi_{B}$ is experimentally extracted as described in \cite{pierret1996semiconductor}, since the Schottky-Mott rule does not take into account the presence of charged impurities at the metal-semiconductor interface. Even so, a similar conclusion can be reached because of the symmetric doping profile induced by the TiN-graphene-TiN configuration.

The carrier temperature profile across the graphene sheet is given by solving the heat transport equation \cite{acshotcarrier, High-Responsivity, asymmetric}:

\begin{equation}
    -\kappa \dfrac{\partial^{2}T_{c}}{\partial x^{2}} + \gamma C (T_{c} - T_{0}) = A_{G} I(x)
\end{equation}

Where $\gamma$ is the carrier cooling rate, $C$ is the carrier heat capacity, $A_{G}$ the effective optical absorption of graphene (see Methods), and $I(x)$ is the intensity profile of the excitation waveguide mode (see section 4 of the supporting information). The product of the carrier temperature profile and the Seebeck coefficient is later integrated to find out the induced PTE voltage, according to Eq. \ref{eq:1}. Finally, the voltage responsivity of the photodetector is calculated by dividing the PTE voltage by the total input optical power in the waveguide. In this work, we plug in parameters taken from experimental reports to the heat transport equation. First, the electrical conductivity ($\sigma$) of graphene is calculated as \cite{acshotcarrier, asymmetric, highresponse}: 
\begin{equation} \label{eq:2}
    \sigma = \sigma_{0}(1+\dfrac{\text{\textmu}^{2}}{\Delta^{2}}) \; , \; \;\sigma_{0} = 5(\dfrac{e^{2}}{h})
\end{equation}

Where $\sigma_{0}$ is the minimum conductivity taken from \cite{asymmetric}, $h$ is Planck's constant, and $\Delta$ is the minimum conductivity plateau; $\Delta \approx 55\,$meV for graphene-on-SiO$_{2}$ \cite{asymmetric, plateau}. The thermal conductivity ($\kappa$) is related to the electrical conductivity ($\sigma$) through the Wiedemann-Franz Law:

\begin{equation} \label{eq:3}
    \kappa = \dfrac{\pi^{2}k_{B}^{2}T}{3e^{2}}\sigma
\end{equation}

The carrier cooling rate ($\gamma$) can be expressed as \cite{cooling, asymmetric}:

\begin{equation} \label{eq:4}
    \gamma = b \: (T+\dfrac{T_{*}^{2}}{T})
\end{equation}

\begin{equation} \label{eq:5}
     b = 2.2 \: \dfrac{g^{2} \varrho k_{B}}{\hbar k_{F}\ell} \; , \; \; \; T_{*} = T_{BG} \sqrt{0.43k_{F}\ell} 
\end{equation}

\begin{equation} \label{eq:6}
    g = \dfrac{D}{\sqrt{2 \rho s^{2}}} \; , \; \; \; \varrho = \dfrac{2\text{\textmu}}{\pi \hbar^{2} v_{F}^2} \; , \; \; \; k_{F}\ell = \dfrac{\pi \hbar \sigma}{e^{2}} \; , \; \; \; T_{BG} = \dfrac{s\hbar k_{F}}{k_{B}}
\end{equation}

The first term in the right hand side of Eq. (\ref{eq:4}) is related to intrinsic scattering processes that dominate at low temperatures, while the second term is related to disorder-assisted scattering which dominates at high temperatures, including room temperature ($T$). $g$ is the electron-phonon coupling constant, $\varrho$ is the density of states, $k_{F}\ell$ is the mean free path, $k_{F}$ is the Fermi wave vector, $T_{BG}$ is the Bloch-Grüneisen temperature, $D=\,$20$\,$eV is the deformation potential constant \cite{disorder-assisted}, $\rho=7.6\times10^{-7}\,$Kg/m$^{2}$ is the mass density of graphene, and $s = 2\times10^{4} \, $m/s is the speed of longitudinal acoustic phonons \cite{grapheneBook}. The carrier heat capacity ($C$) is given by \cite{elsevier}:

\begin{equation}
    C = \dfrac{\pi^{2}k_{B}^2T}{3}\varrho
\end{equation}

The carrier cooling length ($\xi$) is related to $\kappa$, $\gamma$, and $C$ by the following relation:

\begin{equation}
    \xi = \sqrt{\dfrac{\kappa}{\gamma C}}
\end{equation}

Now we have the ingredients required for solving the heat transport equation. The carrier temperature profile is calculated using the analytical solution to the heat transport equation \cite{cooling, jacek}:

\begin{equation}
    \Delta T = T_{c}(x) - T_{0} = \dfrac{\xi sinh ((x_{0} - |x|)/ \xi)}{2 cosh(x_{0}/\xi)} \left( \dfrac{A_{G} I(x)}{\kappa} \right) 
\end{equation}

Where $x_{0}$ is the distance from the peak of $I(x)$ to the side electrode. The carrier temperature is multiplied by the Seebeck coefficient, which is given by the Mott Formula:

\begin{equation}
    S = - \dfrac{\pi^{2} k_{B}^{2} T}{3e} \dfrac{1}{\sigma} \dfrac{d\sigma}{d\mu}
\end{equation}

The photocurrent is calculated by dividing the resultant PTE voltage by the resistance of the graphene sheet ($R_{G}$):

\begin{equation}
    R_{G} = \dfrac{w}{L}\sigma^{-1}(x) 
\end{equation}

Where $w$ is the electrodes spacing, and $L$ is the photodetector length. Finally, the current responsivity is calculated by dividing the photocurrent by the total input optical power in the waveguide.

\section{Methods}
\label{sec:methods}

The optimization procedure, presented in supplementary section 1, was carried out for the graphene-on-SiO$_{2}$ waveguide photodetector. The refractive index of TiN was taken as $n = 2.54 + 7.84i$ at $\lambda = 1550\,$nm \cite{embedded, gosciniak_justice_khan_corbett_2016} (see supplementary section 3). Simulations were conducted using Lumerical, where graphene is modeled as a 2D material with a surface optical conductivity ($\Tilde{\sigma}$) given by \cite{grapheneModel, Falkovsky_2008}:

\begin{equation} \label{eq:1st}
    \Tilde{\sigma}(\omega, \Gamma, \text{\textmu}, T) = \Tilde{\sigma}_{intra}(\omega, \Gamma, \text{\textmu}, T) + \Tilde{\sigma}_{inter}(\omega, \Gamma, \text{\textmu}, T)
\end{equation}

\begin{equation} \label{eq:2nd}
     \Tilde{\sigma}_{intra}(\omega, \Gamma, \text{\textmu}, T) = \dfrac{-je^2}{\pi\hbar^2(\omega+j2\Gamma)}\int_{0}^{\infty}E\,(\dfrac{\partial f(E)}{\partial E} - \dfrac{\partial f(-E)}{\partial E}) \: dE
\end{equation}

\begin{equation} \label{eq:3rd}
     \Tilde{\sigma}_{inter}(\omega, \Gamma, \text{\textmu}, T) = \dfrac{-je^2(\omega+j2\Gamma)}{\pi\hbar^2}\int_{0}^{\infty}\dfrac{f(-E) - f(E)}{(\omega+j2\Gamma)^2 - 4(E/\hbar)^2} \: dE
\end{equation}

\begin{equation} \label{eq:4th}
     f(E) = (e^{(E-\mu)/k_{B}T}+1)^{-1}
\end{equation}

\begin{figure}
\begin{subfigure}{.49\textwidth}
  \centering
  \includegraphics[width=1\linewidth]{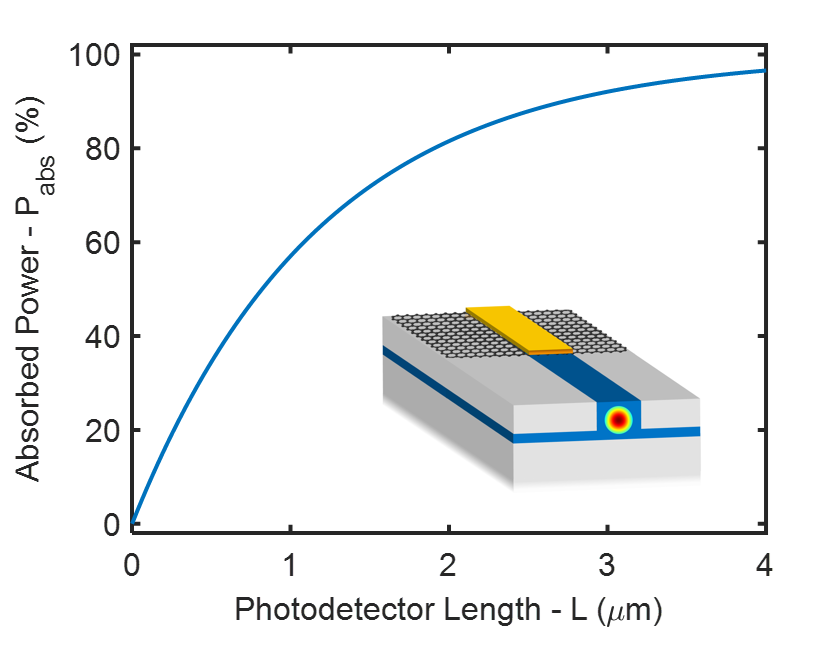}
  \caption{}
  \label{fig:absPow}
\end{subfigure}%
\begin{subfigure}{.49\textwidth}
  \centering
  \includegraphics[width=1\linewidth]{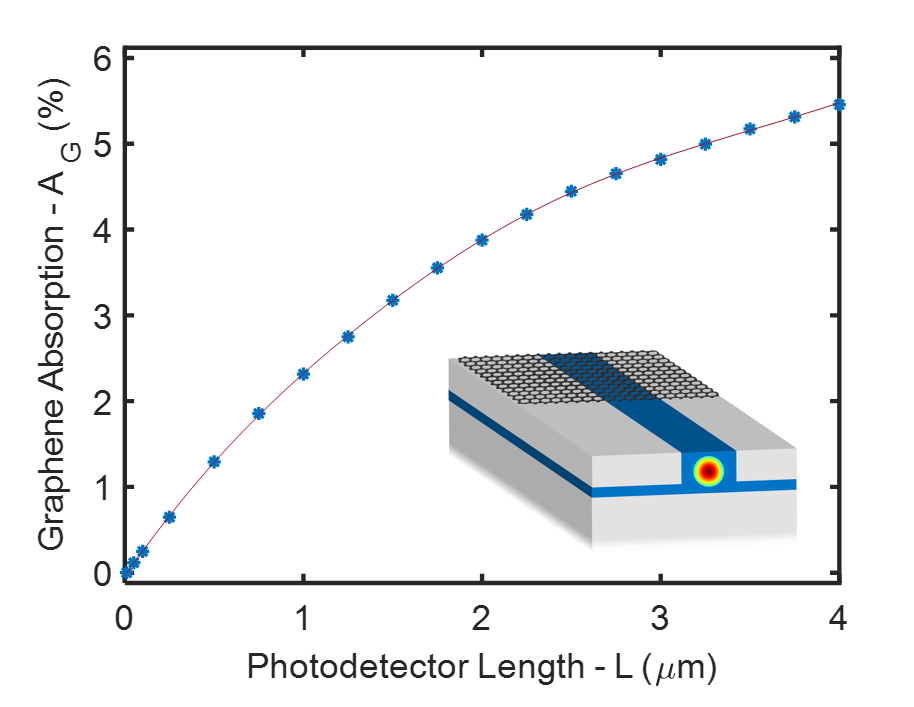}
  \caption{}
  \label{fig:gabs}
\end{subfigure}
\caption{(a) Optical power absorbed as a function of the photodetector length for the optimized waveguide geometry, and (b) Effective absorption of graphene as a function of the photodetector length for the imported optimum mode at $\lambda=1550\,$nm. Data was extracted from power monitors in Lumerical FDTD for the structure shown in inset.}
\label{fig:simulation}
\end{figure}

$\Tilde{\sigma}_{intra}$ and $\Tilde{\sigma}_{inter}$ account for the surface optical conductivity due to intraband and interband transitions, respectively. $\omega$ is the angular frequency of incident photons, $\Gamma$ is the scattering rate of graphene, $T$ is the operation temperature, $e$ is the electron charge, $\hbar$ is the reduced Planck constant, $f(E)$ is the Fermi-Dirac distribution, and $k_{B}$ is the Boltzmann constant. The surface electric permittivity ($\Tilde{\epsilon}$) and the surface electric susceptibility ($\Tilde{\chi}$) are related to $\Tilde{\sigma}$ through the following relation \cite{grapheneBook}:

\begin{equation}
    \Tilde{\epsilon} = \epsilon_{0} \Tilde{\chi} + j\dfrac{\Tilde{\sigma}}{\omega}
\end{equation}

In general, graphene samples are unintentionally doped when placed on a substrate, which results in a chemical potential in the range of 0.1--$\,$0.2$\:$eV \cite{all-optical, doi:10.1021/nn800354m}. In this work, we model graphene with a 0.15$\:$eV chemical potential and 1$\:$ps scattering time. The scattering time is related to the scattering rate by $\tau = 1/2\Gamma$. An incident photon has a $\sim\,$0.8$\:$eV energy at $\lambda=1550\:$nm; this photon induces an interband transition when absorbed by graphene since $\hbar\omega>\,$2\textmu$ $ (see supplementary section 2). The optical absorption of graphene is dominated by interband transitions at telecom wavelengths for the aforementioned chemical potential range. Therefore, variations in the scattering rate, which is related to intraband transitions, will have a negligible effect on the optical absorption of graphene for 0.1$\:$eV $\leq \text{\textmu} \leq$ 0.2$\:$eV. Our simulations revealed that the optical absorption of graphene was similar for 100$\:$fs, 500$\:$fs and 1000$\:$fs scattering times, as explained in supplementary section 2. On the other hand, chemical potential variations have a significant impact on the device performance, where the responsivity degrades for large \text{\textmu} values. In our case, the responsivity reduction associated with larger $\text{\textmu}$ is mainly determined by the drop in $T_{c}$ and $S$ at higher carrier densities, as is explained in section \ref{sec:results}.

The computed propagation loss is 3.67$\:$dB/\textmu m for the TM-mode presented in Fig S5 of the supporting information. Here the propagation loss ($\alpha$) is defined as \cite{propagationLoss}:

\begin{equation}
    \alpha = -20 \: \textrm{log}_{10} \: (E_{f}/E_{i})
\end{equation}

Where $E_{i}$ and $E_{f}$ represent the electric field intensity before and after propagating through the photodetector waveguide, respectively. The absorbed power ($P_{abs}$) in the photodetector waveguide can be calculated using the Beer-Lambert law:

\begin{equation}\label{eq:beer}
    P_{abs}(L) = 1 - 10^{-2(\alpha/20)L}
\end{equation}

Fig \ref{fig:absPow} shows $P_{abs}$ as a function of the photodetector length for $\alpha=3.6\:$dB/\textmu m. $P_{abs}= 95\%$ for $L=3.5\:$\textmu m, as was given in supplementary section 1. To find out the effective absorption of graphene ($A_{G}$), we extract the computed mode shown in supplementary Fig S5, and then import it into a Lumerical FDTD simulation of the same device structure shown in Fig \ref{fig:sfig2}, but without the TiN stripe. The presence of the TiN stripe introduces a plasmonic near-field enhancement of the EM field at the graphene sheet plane. In order to find out the power absorbed solely by graphene, we follow the aforementioned approach, where the only absorbing material in the Lumerical FDTD simulation is graphene, which "sees" the plasmonically enhanced EM field at the graphene sheet plane due to the presence of the TiN stripe in the imported mode. Fig \ref{fig:gabs} plots $A_{G}$ as a function of the photodetector length for the imported plasmonic TM-mode with TiN, where $A_{G} = 5.2$\% for $L=3.5\:$\textmu m. We find out that the effective absorption of graphene is \textgreater2$\times$ larger purely because of the presence of the TiN stripe, where the absorption of the graphene-silicon waveguide without the TiN stripe is only 2.4\% for $L=3.5\:$\textmu m (see supplementary section 1). Hence, the effectively enhanced absorption is a consequence of the plasmonic near-field enhancement of the TiN stripe, which the TM-mode experiences as it propagates in the photodetector section. 

\section{Results and Discussion}
\label{sec:results}
\subsection{Carrier Cooling and Thermoelectric Performance}
\label{superior}

\begin{figure}
\begin{subfigure}{.49\textwidth}
  \centering
  \includegraphics[width=1\linewidth]{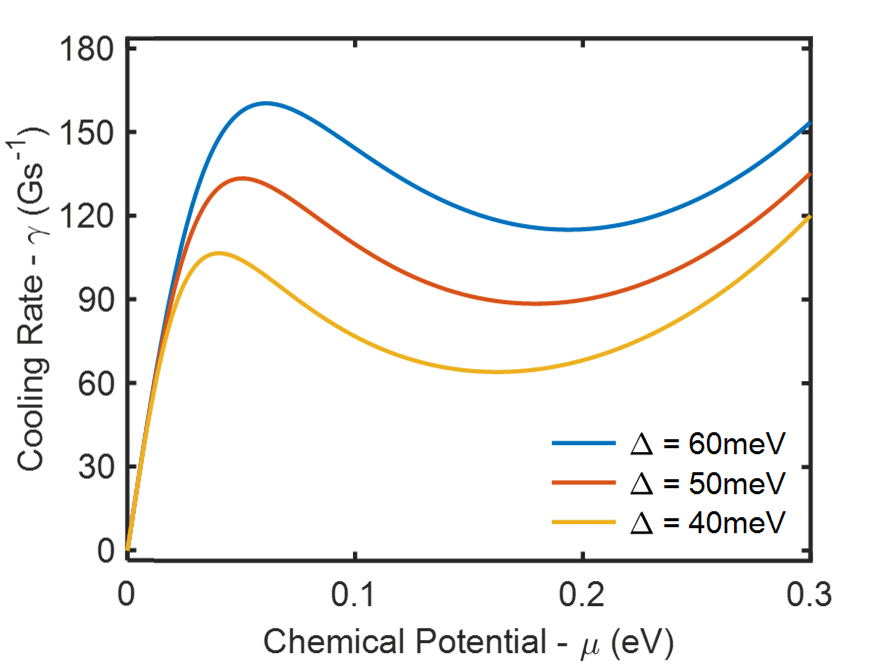}
  \label{fig:a}
\end{subfigure}%
\begin{subfigure}{.49\textwidth}
  \centering
  \includegraphics[width=1\linewidth]{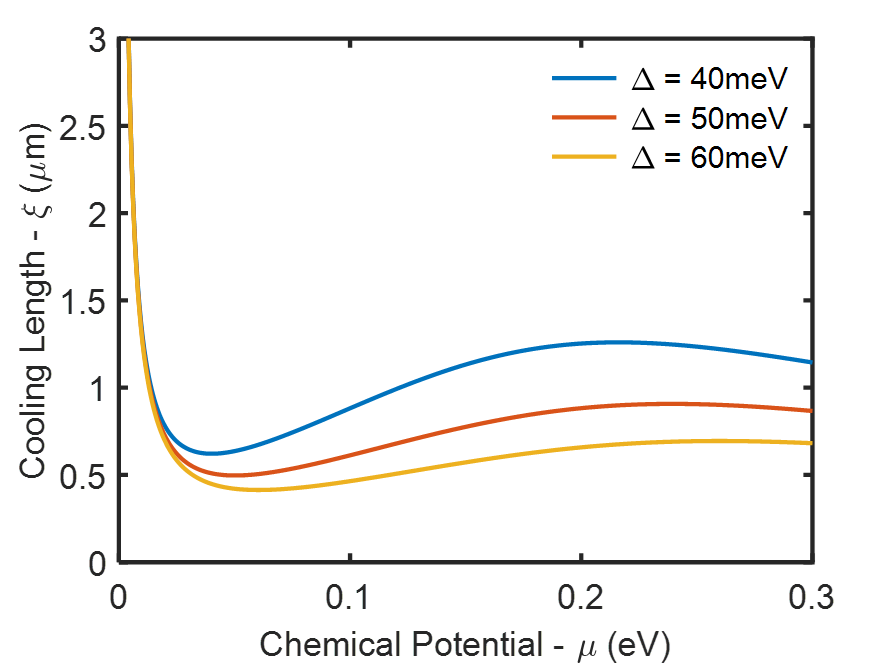}
  \label{fig:b}
\end{subfigure}
\caption{Carrier cooling rate and cooling length as a function of chemical potential for $\Delta=60\,$meV, $\Delta=50\,$meV, and $\Delta=40\,$meV.}
\label{fig:result1}
\end{figure}

Fig \ref{fig:result1} shows the carrier cooling rate and cooling length as a function of chemical potential. It is noted that carriers cool down faster in graphene with higher disorder, i.e. higher $\Delta$, resulting in a faster operation speed. The highest demonstrated bandwidth for a PTE graphene on-chip detector is 67$\:$GHz \cite{2020ultrafast}, which is a setup-limited bandwidth. A bandwidth of $\sim \,$110$\,$GHz has been demonstrated for graphene photodetectors operating based on the photovoltaic and bolometric effects \cite{plasmonic4, plasmonic1}. A bandwidth exceeding 100$\,$GHz is not inconceivable for PTE graphene photodetectors considering the ultrafast cooling dynamics of photoexcited hot carriers in graphene. Experimentally, the cooling time of graphene is on the order of a few picoseconds \cite{bao2009atomic, nano}, corresponding to a cooling rate in hundreds of Gs$^{-1}$. Furthermore, it is experimentally reported that highly-disordered graphene sheets exhibit faster carrier cooling dynamics than less disordered sheets \cite{dawlaty2008measurement}, which is in agreement with our conclusions. The cooling length goes like $\xi\sim1/\sqrt{\gamma}$, hence the higher the cooling rate, the shorter is the cooling length. 

\begin{figure}
\begin{subfigure}{.49\textwidth}
  \centering
  \includegraphics[width=1\linewidth]{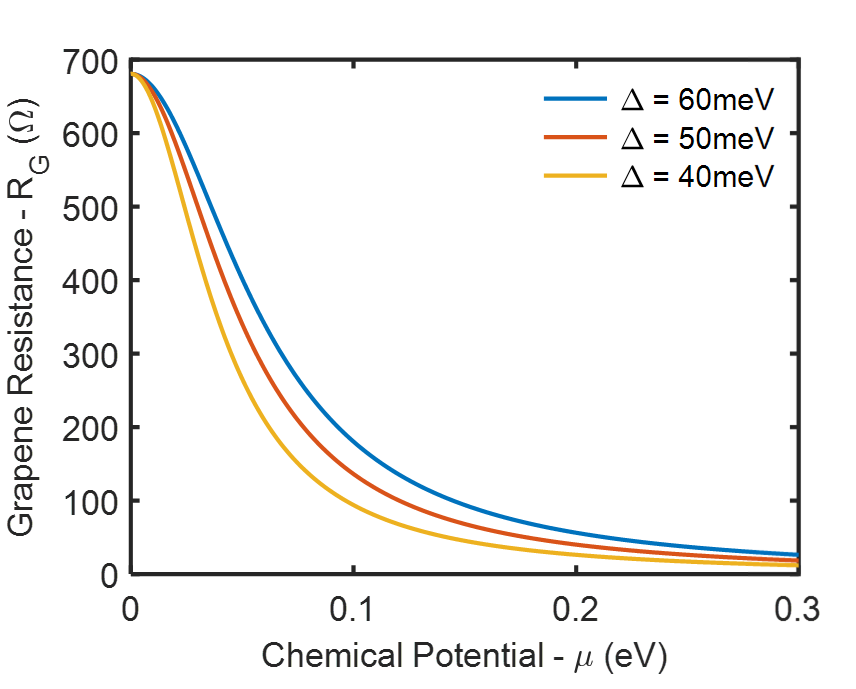}
  \label{resistance}
\end{subfigure}
\begin{subfigure}{.49\textwidth}
  \centering
  \includegraphics[width=1\linewidth]{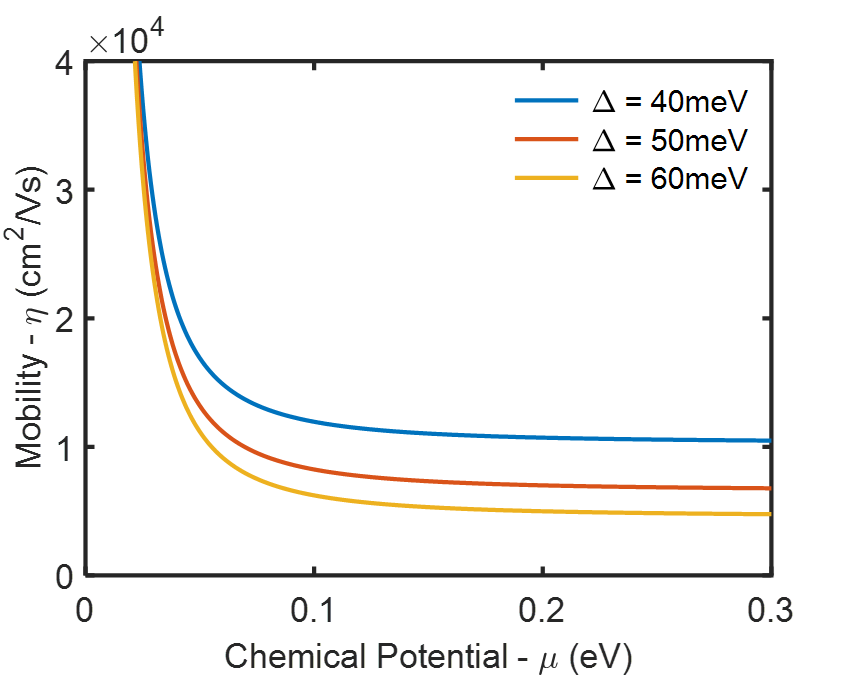}
  \label{mobility}
\end{subfigure}
\caption{Graphene sheet resistance and mobility as a function of the chemical potential for $\Delta=60\,$meV, $\Delta=50\,$meV, and $\Delta=40\,$meV.}
\label{fig:resistanceMu}  
\end{figure}

\begin{figure}
\begin{subfigure}{.49\textwidth}
  \centering
  \includegraphics[width=1\linewidth]{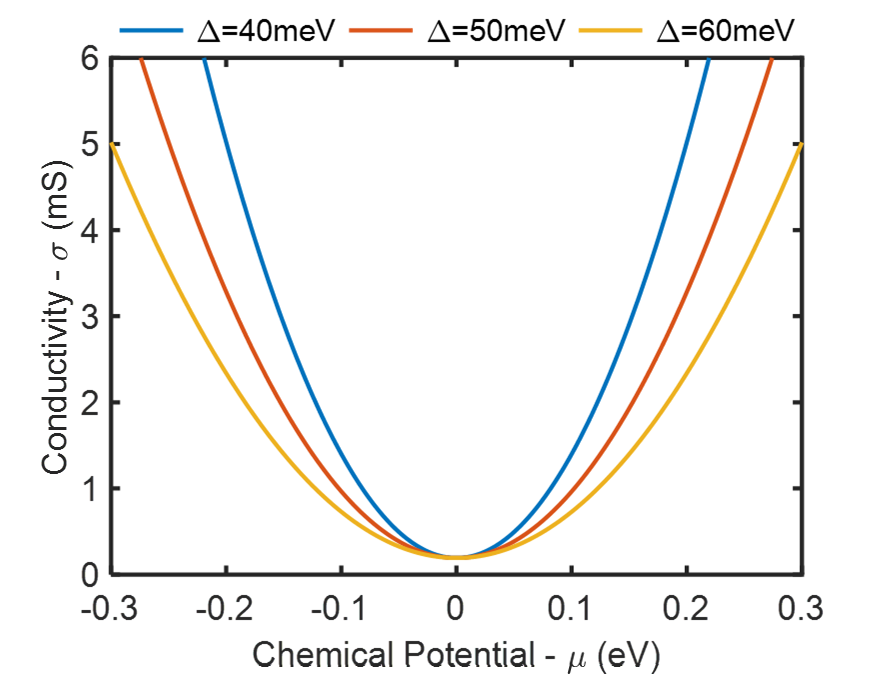}
  \caption{}
  \label{fig:conduct1}
\end{subfigure}%
\begin{subfigure}{.49\textwidth}
  \centering
  \includegraphics[width=1\linewidth]{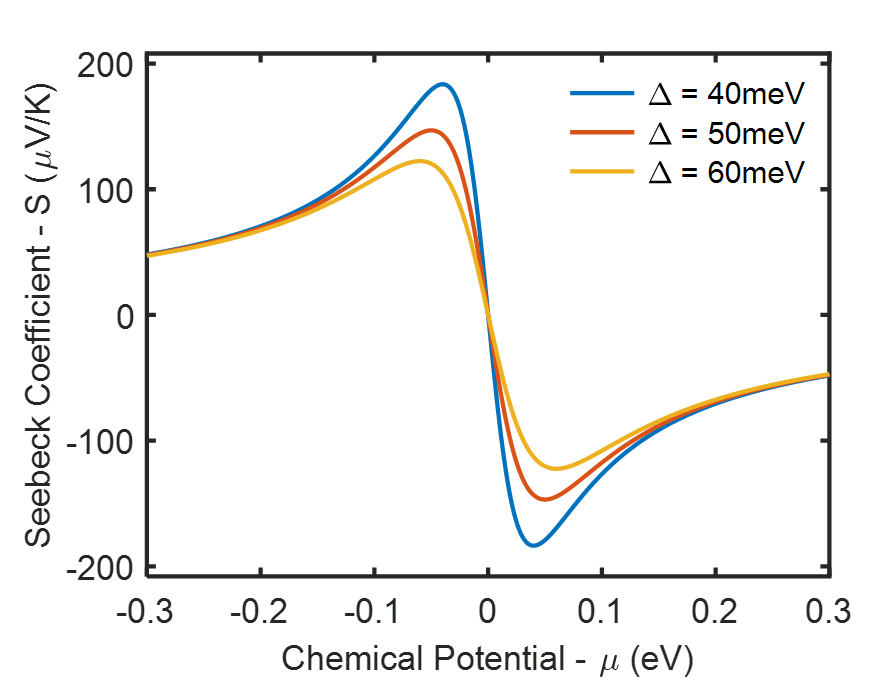}
  \caption{}
  \label{fig:seebeck2}
\end{subfigure}
\caption{Electrical conductivity of graphene and Seebeck coefficient as a function of the chemical potential for $\Delta=40\,$meV, $\Delta=50\,$meV, and $\Delta=60\,$meV.}
\end{figure}

As previously mentioned, carrier cooling mechanisms are dominated by disorder-assisted scattering at room temperature. Disorder can come in the form of ripples, charged impurities, or strain fluctuations \cite{clean1, charged-impurity}. Low-disorder graphene has a lower resistance and higher mobility than high-disorder graphene, as shown in Fig \ref{fig:resistanceMu}, where the mobility ($\eta$) is taken as the Drude mobility $\eta = \sigma/en$, and $n = \mu^2 / \pi \hbar^2 v_{F}^2$ is the density of carriers in graphene \cite{correlated}. High-mobility carriers in low-disorder graphene experience less scattering, resulting in a slower cooling rate for photoexcited hot carriers. On the other hand, in high-disorder graphene, the mobility is relatively low, and carriers are more likely to scatter and cool down rapidly as a result. Disorder is accounted for through the minimum conductivity plateau ($\Delta$). Higher disorder is manifested in the form of a wider charge neutrality region in the conductivity plot of graphene \cite{charged-impurity}, as shown in Fig. \ref{fig:conduct1}. To extract the $\Delta$ parameter, the conductivity plot of graphene is experimentally measured, and then analytically fitted with the appropriate $\Delta$ value. This was the approach reported by other groups \cite{asymmetric, plasmon-induced}. The electron mobility in graphene can be higher than the hole mobility \cite{plateau}. In case further modeling of this phenomenon is intended, two distinct $\Delta$ values should be adopted, where the electron-related $\Delta$ would be lower than the hole-related $\Delta$, since the electron mobility is higher. Depending on whether the graphene sheet is p- or n-doped, the appropriate $\Delta$ value should be used.

\begin{figure}
\begin{subfigure}{.49\textwidth}
  \centering
  \includegraphics[width=1\linewidth]{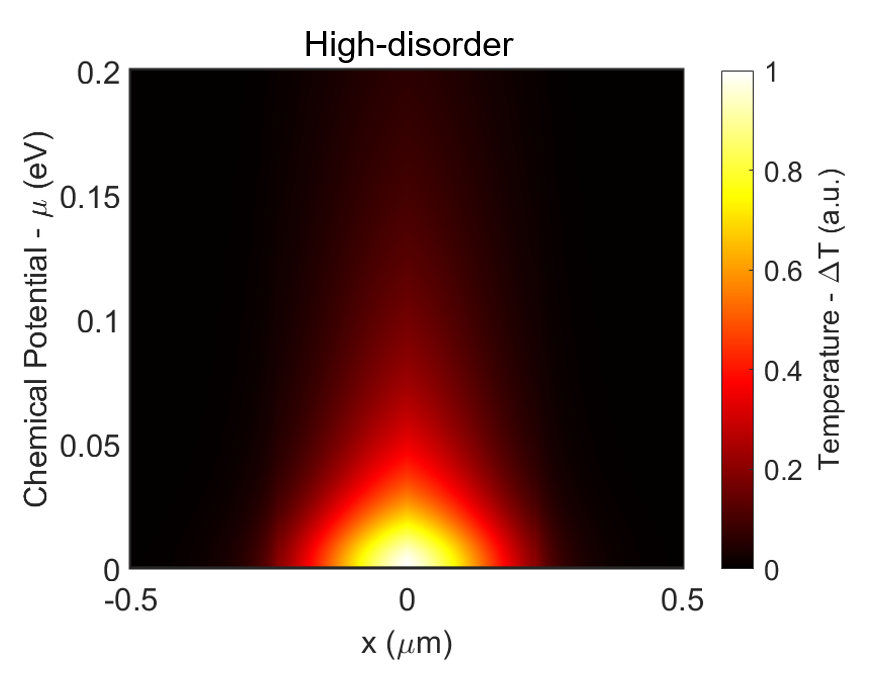}
  \label{fig:oxide}
\end{subfigure}%
\begin{subfigure}{.49\textwidth}
  \centering
  \includegraphics[width=1\linewidth]{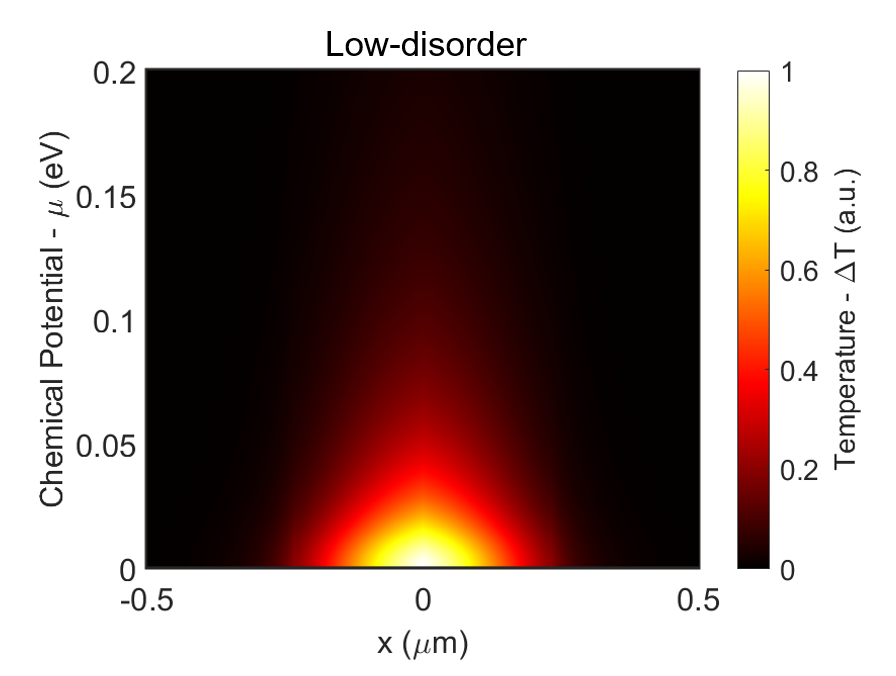}
  \label{fig:boron}
\end{subfigure}
\caption{Normalized temperature profiles for a varying chemical potential for the high-disorder ($\Delta=60\,$meV) and low-disorder ($\Delta=40\,$meV) graphene sheet.}
\label{fig:Tc}
\end{figure}

The normalized $T_{c}$ profiles of the high-disorder ($\Delta=60\,$meV) and the low-disorder ($\Delta=40\,$meV) graphene photodetectors are shown in Fig \ref{fig:Tc}. The maximum $T_{c}$ is located at the waveguide center, and decays as one approaches the waveguide sides. Therefore, a strong $T_{c}$ gradient is present across the graphene sheet, resulting in the generation of a PTE voltage according to Eq. (\ref{eq:1}). It is noted that $T_{c}$ drops with rising \textmu, which is expected since the PTE effect is more dominant in the low $\text{\textmu}$ regime, as was explained in section 3. It is observed that the carrier temperature is slightly more concentrated at low chemical potentials for the low-disorder sheet. A similar trend is seen in the Seebeck coefficient ($S$) plots shown in Fig \ref{fig:seebeck2}, where $S$ is higher for low-disorder sheets at low chemical potentials. The Seebeck coefficient, and hence the thermoelectric performance, is enhanced in low-disorder graphene \cite{zhou2015spin}. 

Photovoltaic graphene photodetectors function based on the flow of photoexcited carriers in graphene, which makes such photodetectors carrier transit-time limited \cite{gao}. On the other hand, PTE graphene photodetectors operate based on the generation of a thermoelectric voltage induced by photoexcited hot carriers in graphene. The operation speed of PTE graphene photodetectors is limited by the carrier cooling rate \cite{nano, jacek, 2020ultrafast}, where carriers can re-participate in the photodetection process after adequately cooling down. Interestingly, a high mobility graphene sheet is favorable for photovoltaic photodetectors, since a low mobility sheet limits the speed of carrier flow across graphene, while a low mobility sheet is preferable for PTE graphene photodetectors, as long as the device bandwidth is the only concern. On the other hand, the Seebeck effect, and consequently the thermoelectric performance is degraded at higher disorder \cite{zhou2015spin}, resulting in a responsivity-bandwidth trade-off for graphene-on-SiO$_2$ PTE detectors that is dictated by the disorder strength.

\subsection{Photoresponsivity and Noise Performance}

\begin{figure}
\begin{subfigure}{.49\textwidth}
  \centering
  \includegraphics[width=1\linewidth]{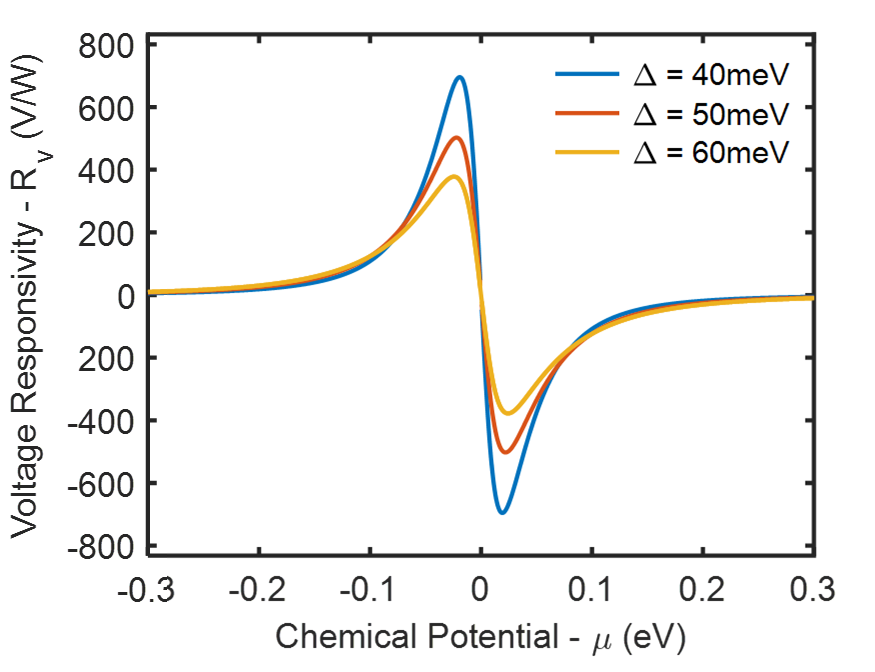}
\end{subfigure}
\begin{subfigure}{.49\textwidth}
  \centering
  \includegraphics[width=1\linewidth]{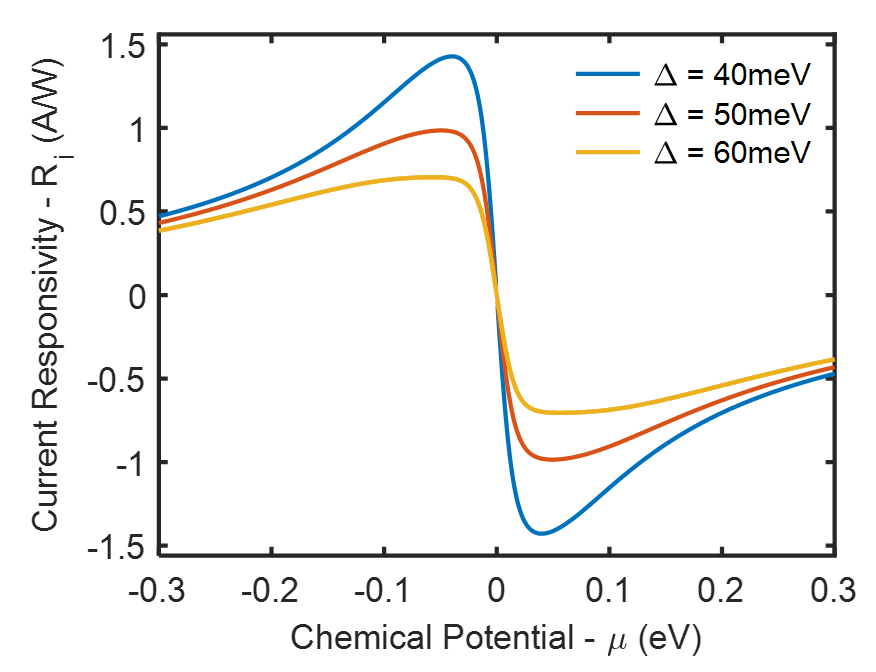}
\end{subfigure}
\caption{Voltage and current responsivity of the photodetector as a function of the chemical potential for $\Delta=40\,$meV, $\Delta=50\,$meV, and $\Delta=60\,$meV.}
\label{fig:responsivity}
\end{figure}

The voltage responsivity ($R_{v}$) and the current responsivity ($R_{i}$) of the photodetector are shown in Fig \ref{fig:responsivity}. The enhanced thermoelectric performance of low-disorder graphene augments the detector's photoresponse. The voltage responsivity exceeds 350$\,$V/W and 650$\,$V/W for $\Delta=60\,$meV and $\Delta=40\,$meV, respectively. It is observed that the responsivity strongly depends on the chemical potential, which is related to the dependence of $T_c$ and $S$ on $\text{\textmu}$. The device performance is optimal when $0< \, \text{\textmu} \, <0.1\:$eV, where the Seebeck coefficient and the carrier cooling rate are both high. Consequently, chemical potential tuning is required to acquire the optimum responsivity. The chemical potential of graphene on a substrate can be tuned by thermal annealing \cite{disorder-assisted, doi:10.1063/1.3599708, doi:10.1021/nn800354m}. The current responsivity is as high as 1.4$\,$A/W (1.1$\,$A/W external) for $\Delta=40\,$meV, and 0.7$\,$A/W (0.6$\,$A/W external) for $\Delta=60\,$meV. 

The highest demonstrated responsivity of a waveguide-integrated graphene-based photodetector is 0.67$\,$A/W at 0.5$\,$V bias, for a $5\,\text{\textmu}$m long device \cite{ma2020compact}. Using high-mobility graphene ($\Delta=40\,$meV), the herein proposed device can deliver $>$$1.5\times$ that responsivity (1.1$\,$A/W external) for a smaller footprint (3.5$\text{\textmu}$m), and is the most compact graphene-based waveguide photodetector reported up-to-date. Moreover, this device operates at the zero-bias condition, while offering CMOS-compatibility and ultra-high-speed beyond 100$\,$GHz. Theoretical responsivities up to 1000$\:$A/W have been reported in \cite{jacek} for a waveguide-integrated graphene photodetector, which may be overestimated considering the highest responsivity demonstrated so far, namely 0.67$\:$A/W. 

\begin{figure}
\begin{subfigure}{.49\textwidth}
  \centering
  \includegraphics[width=0.98\linewidth]{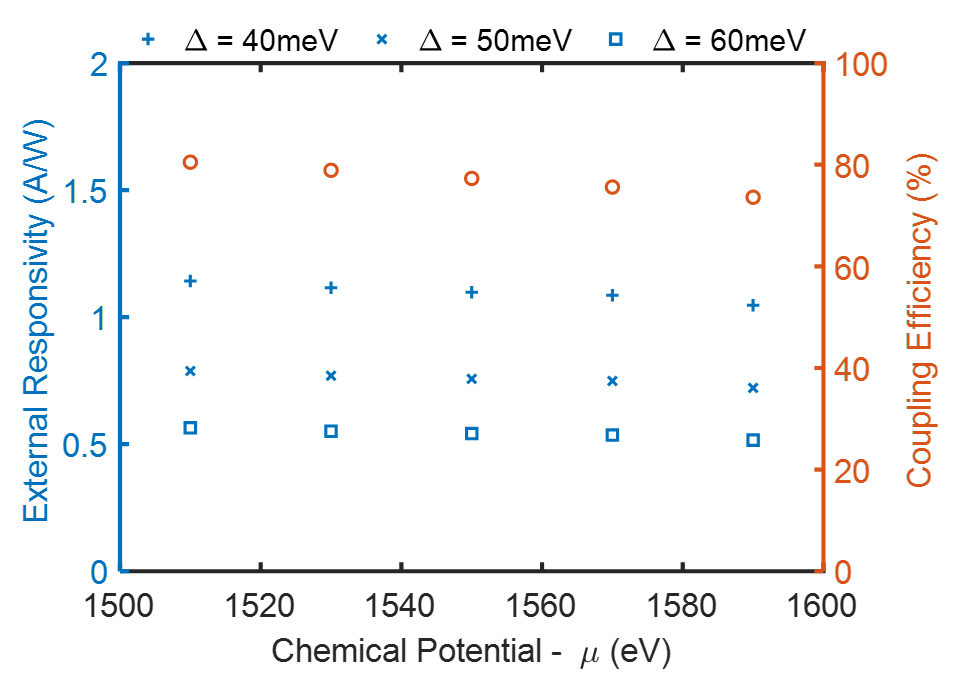}
  \caption{}
  \label{fig:wavelength}
\end{subfigure}
\begin{subfigure}{.49\textwidth}
  \centering
  \includegraphics[width=0.98\linewidth]{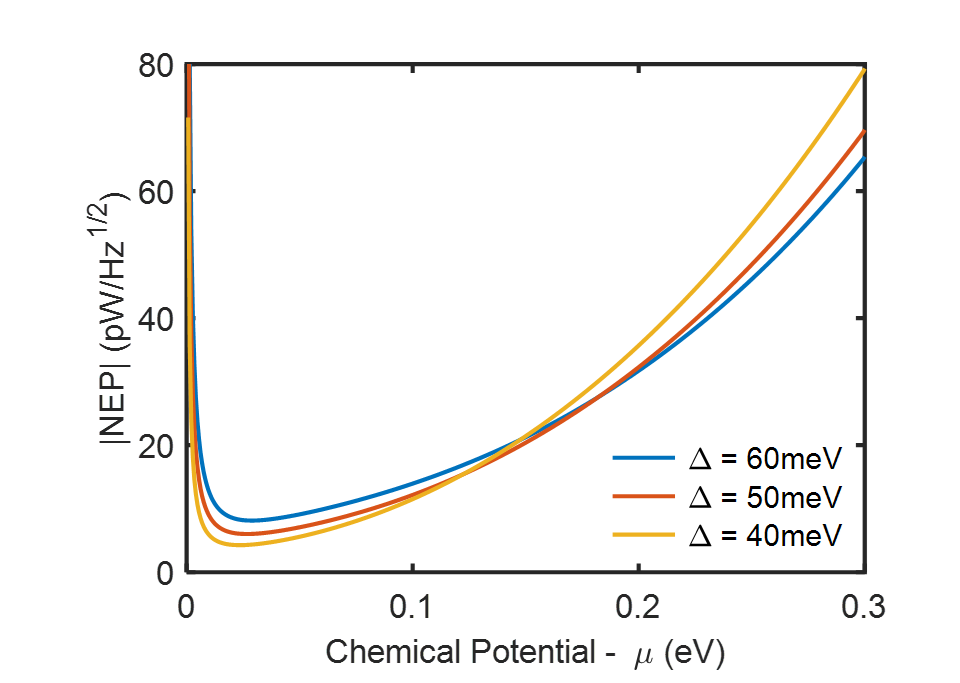}
  \caption{}
  \label{fig:NEP}
\end{subfigure}
\caption{(a) Maximum external current responsivity and coupling efficiency as a function of wavelength, and (b) Noise equivalent power (NEP) as a function of chemical potential, for $\Delta=40\,$meV, $\Delta=50\,$meV, and $\Delta=60\,$meV.}
\label{fig:end}  
\end{figure}

The photodetector exhibits a nearly flat photoresponse across the telecom C-band, as illustrated by the blue markers in Fig \ref{fig:wavelength}. Here we consider the maximum external responsivity of the photodetectors for comparison, where the external responsivity is defined as the product of the responsivity with the coupling efficiency. It is observed that the responsivity slightly increases at shorter wavelengths following the trend of the coupling efficiency, which is represented by the red markers in Fig \ref{fig:wavelength}. Certainly, the graphene absorption of the propagating mode increases at longer wavelengths, since the propagating mode becomes more confined in the photodetector waveguide, and the graphene sheet absorbs more of the propagating mode as a result (see supplementary section 3). However, at long wavelengths in this band, the coupling efficiency reduces at a higher rate than the increasing rate of the graphene absorption, which explains the results that we obtain in this study. 

Further improvement of the device performance can be achieved by placing graphene on a hexagonal boron nitride (hBN) substrate, or encapsulating graphene between two hBN layers. The Seebeck coefficient of graphene is enhanced by $\sim$$2\times$ when placed on a hBN substrate \cite{plasmon-induced}. This is due to the charge screening role of the hBN spacer layer, which mitigates the influence of the substrate impurities \cite{Duan14272}. In addition, the carrier cooling dynamics in graphene-hBN Van der Waals (vdW) heterostructures is dominated by hot carrier coupling to hyperbolic phonon polaritons, not disorder-assisted scattering \cite{tielrooij2018out, principi2017super, huang2020ultra}. Experimentally, graphene-on-hBN demonstrated faster cooling dynamics than graphene-on-SiO$_2$ \cite{tielrooij2018out, golla2017ultrafast}. Therefore, the cooling pathway introduced by hyperbolic phonon polaritons is more efficient than disorder-assisted scattering, which opens up the possibility for faster optoelectronic devices that are based on graphene-hBN vdW heterostructures.

A transimpedance amplifier is an electronic circuit that converts an input electric current into voltage, and is typically employed in receiver circuits to convert the photodetector current into a voltage reading \cite{jawdat}. The proposed device can be operated without a transimpedance amplifier \cite{2D}, since a photovoltage is already generated across the graphene sheet, and can be collected through the two TiN films as was shown in Fig. \ref{fig:responsivity}. Moreover, this device does not consume energy, has a zero dark current, zero flicker ($1/f$) noise, and zero shot noise, because of its zero-bias operation \cite{8510130, yurgen, perepelitsa2006johnson, gray2001analysis, BAHREYNI2009129, 2020ultrafast}. Therefore, its noise equivalent power (NEP) is determined by the Johnson-Nyquist (thermal) noise:

\begin{equation} \label{eq:nep}
    NEP = \dfrac{V_{th}}{R_{v}} = \dfrac{\sqrt{4k_{B}TR_{G}}}{R_{v}}
\end{equation} 

Where $V_{th}$ is the variance of the thermal noise voltage per 1 Hz of bandwidth. The NEP is defined as the input signal power that results in an SNR ratio of 1 in a 1 Hz output bandwidth. A low NEP value corresponds to a lower noise floor and  a more sensitive detector \cite{NEP, Leclercq2007DiscussionAN}. Therefore, a low NEP value is an attractive feature for photodetector devices. Fig \ref{fig:NEP} shows the NEP as a function of \textmu, for $\Delta=40\,$meV, $\Delta=50\,$meV, and $\Delta=60\,$meV. It is noted that the NEP for the low-disorder photodetectors is initially low at low chemical potentials. That is attributed to the remarkably high $R_{v}$ and small $R_{G}$ of the low-disorder graphene photodetector in that regime. Nevertheless, the NEP becomes comparatively lower for the high-disorder graphene photodetector at high chemical potentials, where the voltage responsivity of the low-disorder graphene photodetector is relatively low. The voltage responsivity of the the high-disorder graphene photodetector becomes slightly higher than that of the low-disorder graphene photodetector at large $\text{\textmu}$ values, as was illustrated in Fig \ref{fig:responsivity}. The resistance of low-disorder graphene is lower than that of high-disorder graphene at all chemical potentials, as was shown in Fig \ref{fig:resistanceMu}; this is supposed to give a lower NEP for the low-disorder graphene at large $\text{\textmu}$ values, but the graphene sheet resistance term ($R_{G}$) is under the square root in Eq. \ref{eq:nep}, making the $R_{v}$ term more effective in determining the NEP. 

Considering that the photodetectors are operating in the optimal chemical potential range, $0<\,$\textmu$\,<0.1\:$eV, the corresponding NEP values are nearly $5<\,|$\text{NEP}$|\,<20\:$pW/$\sqrt{\text{Hz}}$ for both photodetectors, where we deliberately exclude the NEP values at $\text{\textmu}=0$, in addition to NEP values in its very close proximity, since the chemical potential cannot be exactly zero, as was explained in section \ref{sec:methods}. These NEP values are an order of magnitude lower than the reported values for other plasmonically-enhanced graphene photodetectors \cite{plasmonic2, plasmonic3}. Therefore, the zero-bias operation of the proposed photodetector eliminates flicker noise, shot noise and dark current, and as a result, the NEP is minimal. However, in an actual measurement setup, noise from the photodetector makes up a fraction of the total noise, where contributions by other components in the system may be dominant, e.g. amplifier noise \cite{plasmonic1}, and taking that into account may invalidate the previous comparison. Nonetheless, the intrinsic NEP introduced by this photodetector is ultra-low (\textless$20\:\text{pW/}\sqrt{\text{Hz}}$).

\subsection{Conclusion}

To sum up, a waveguide-integrated graphene photodetector based on the plasmon-induced photothermoelectric effect is proposed. The on-chip photodetector relies on the plasmonic response of titanium nitride (TiN), which is a CMOS-compatible metal nitride that exhibits plasmonic properties similar to gold.  Simulations were carried out to optimize the device geometry, and hot carrier transport theory was applied to study the device performance. We conclude that higher device speeds are possible with high-disorder graphene, while a superior thermoelectric performance is achievable using low-disorder graphene. The effect of a varying chemical potential was studied, and it was found out that the device performance is optimal for $0<\,$\textmu$\,<0.1\:$eV, because the carrier cooling rate and the Seebeck coefficient are both high in this range. Finally, the device responsivity was calculated, where it was found out that low-disorder graphene is more responsive than the high-disorder graphene, with a responsivity as high as 1.4$\,$A/W (1.1$\,$A/W external) for the low-disorder graphene ($\Delta=40\,$meV), and 0.7$\,$A/W (0.6$\,$A/W external) for the high-disorder graphene ($\Delta=60\,$meV) photodetectors, at an ultra-compact length of 3.5$\,$\textmu m. This device has the most compact footprint reported for a waveguide-integrated graphene-based photodetector. Moreover, it exhibits a nearly flat photoresponse across the telecom C-band. In addition, the photodetector has an ultra-low intrinsic NEP (\textless$20\:\text{pW/}\sqrt{\text{Hz}}$) at the optimal chemical potential range, because of its bias-free operation, which also means that it consumes zero energy. We draw out these conclusions after performing our study on plasmonically-enhanced photothermoelectric graphene photodetectors in a waveguide-integrated configuration. Further advances in graphene deposition methods and quality control of samples will be key in realizing graphene-based optoelectronics as a mainstream industrial technology.

\begin{acknowledgement}

Support from the NYUAD Research Grant is gratefully acknowledged.

\end{acknowledgement}

\section*{Suppementary Material}
See Supplement 1 for supporting content.



\bibliography{achemso-demo}

\end{singlespace}

\end{document}


\begin{singlespace}

\section{Design and Optimization}
\label{sec:design}

\begin{figure}[H]
\begin{subfigure}{.5\textwidth}
  \centering
  \includegraphics[width=1\linewidth]{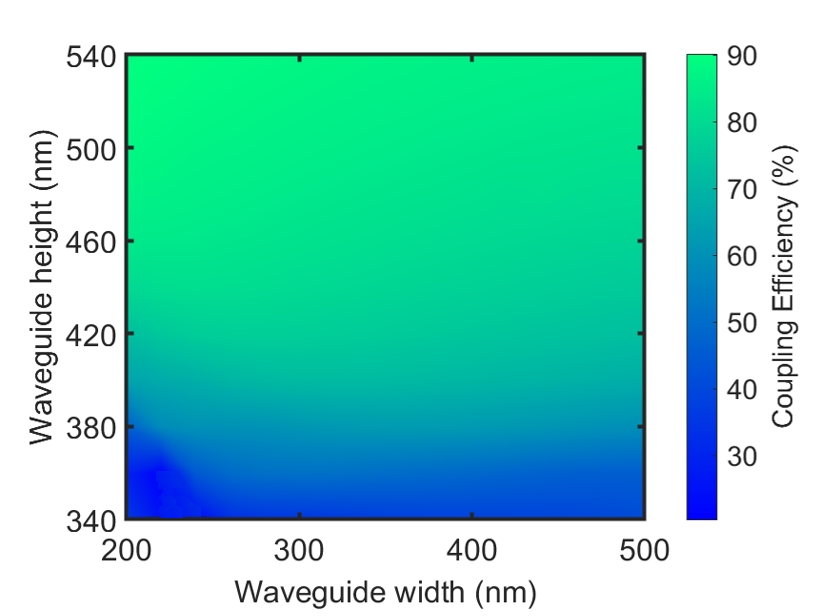}
  \caption{}
  \label{fig:optA}
\end{subfigure}%
\begin{subfigure}{.5\textwidth}
  \centering
  \includegraphics[width=0.99\linewidth]{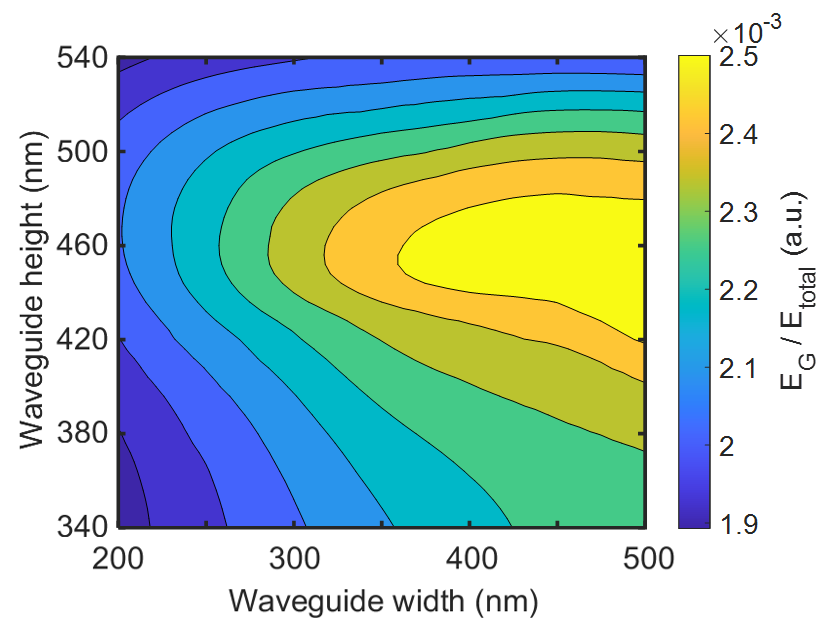}
  \caption{}
  \label{fig:optB}
\end{subfigure}
\caption{(a) Coupling efficiency between the SOI waveguide and the photodetector section, and (b) relative electric field intensity at the TiN-graphene interface as a function of the waveguide height and width at $\lambda=1550\,$nm.}
\label{fig:opt}
\end{figure}

The waveguide geometry is optimized to achieve a high coupling efficiency and a compact footprint. The waveguide height is swept from 340$\,$nm to 540$\,$nm, and its width is swept from 200$\,$nm to 500$\,$nm. The thickness of the silicon slab and the TiN film are initially set to 100$\,$nm and 20$\,$nm, respectively. The coupling efficiency is computed by calculating the power overlap between the fundamental TM-mode of the SOI waveguide and the guided TM-mode that yields the highest coupling efficiency in the photodetector waveguide for each height and width, as we shall explain below. The computation results show that the coupling efficiency increases for thicker waveguides, and is generally less sensitive to width variations, as illustrated in Fig. \ref{fig:optA}.

We quantify the relative electric field intensity ($E$) at the TiN-graphene interface for the photodetector waveguide modes whose coupling efficiencies were computed. A high $E$ field intensity at the TiN-graphene interface is an indication of a strong plasmonic enhancement of the EM field that is present at the graphene sheet plane. Therefore, when optimizing the relative $E$ field intensity, higher responsivities can be obtained for shorter device lengths, since the EM field interaction with the graphene sheet is maximized, as well as the optical absorption of graphene. The relative $E$ field intensity at the TiN-graphene interface is expressed as:

\begin{equation}
    \dfrac{E_{G}}{E_{total}} = \dfrac{\int_{-\infty}^{+\infty} \int_{-\infty}^{+\infty}E(x, y_{g}, z) \;dxdz}{\int_{-\infty}^{+\infty}\int_{-\infty}^{+\infty} \int_{-\infty}^{+\infty}E(x, y, z) \; dxdydz}
\end{equation}

\begin{figure}
  \centering
  \includegraphics[width=0.8\linewidth]{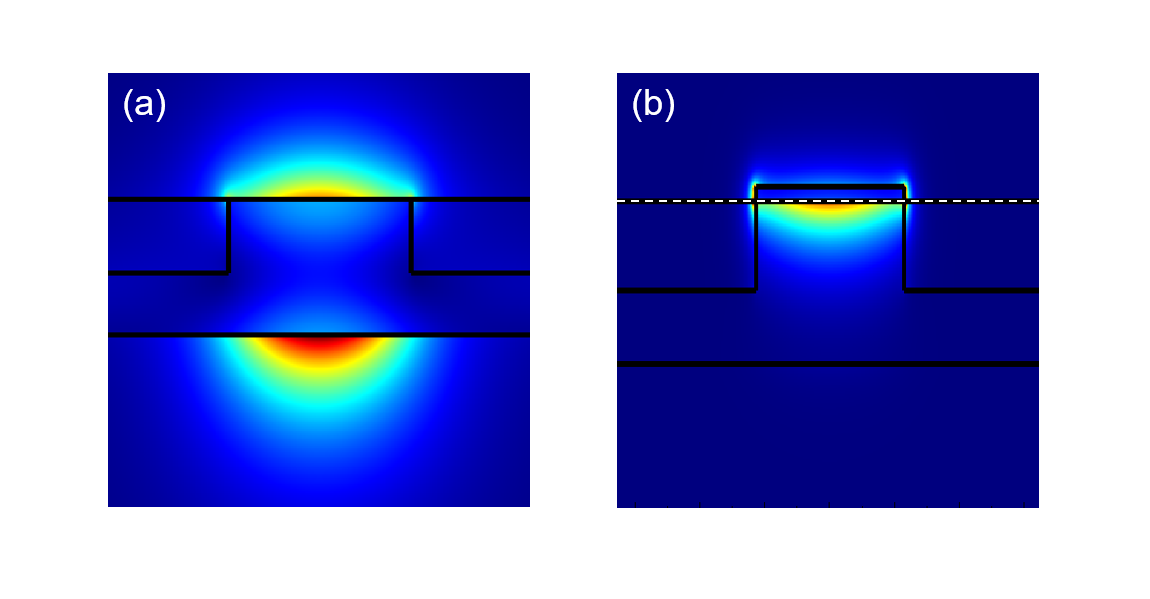}
\caption{(a) Electric field intensity profile of the fundamental TM-mode for a 220$\,$nm$\times$460$\,$nm SOI waveguide. (b) Electric field intensity profile of the fundamental TM-mode for a 220$\,$nm$\times$460$\,$nm SOI waveguide with a graphene sheet and TiN stripe on top of the waveguide at $\lambda=1550\,$nm. The dashed white line represents the graphene sheet plane.}
\label{fig:220example}
\end{figure}

Where $E_{G}$ is the $E$ field intensity at the graphene sheet plane, $y_{g}$ is the height at which the graphene sheet is placed, and $E_{total}$ is the total $E$ field intensity in the waveguide. The relative $E$ field intensity for the aforementioned sweep range is shown in Fig. \ref{fig:optB}. We find out that the maximum relative $E$ field intensity occurs at 460$\,$nm height, which we choose as the optimum height to obtain the most compact footprint. One can also choose a standard height of 500$\,$nm for a reasonable reduction in the relative $E$ field intensity. The computed coupling efficiency is 79\% for a 460$\,$nm thick waveguide, and is 84\% for 500$\,$nm thickness. Based on the optimization results, this photodetector is most suited for applications where III-V lasers are heterogeneously integrated on SOI structures, where thick waveguides are essential to achieve decent coupling efficiencies \cite{8550947, 6697878, sweetSpot}, or other applications where a similar thickness is required. For this device, the coupling efficiency is generally 10$\,$–$\:$20\% for thicknesses below 340$\:$nm, following the trend shown in Fig. \ref{fig:optA}. Therefore, the performance at such small thicknesses was not studied in-depth, since the fraction of optical power coupling to the photodetector is minimal, which makes it detrimental to the external responsivity of the device. For instance, Fig. \ref{fig:220example}a shows the fundamental TM-mode of a 220$\,$nm$\times$460$\,$nm SOI waveguide; 220$\,$nm is the waveguide height, and $460\,$nm is the waveguide width. Most of the $E$ field intensity is concentrated at the top and bottom edges of the silicon waveguide, which yields a high relative $E$ field intensity at the graphene sheet plane. However, the presence of a TiN stripe on top of the waveguide has an immense impact on the fundamental TM-mode, as is shown in Fig. \ref{fig:220example}b. The computed coupling efficiency between the two modes presented in Fig. \ref{fig:220example} is merely 14.3\%. 

\begin{figure}
  \centering
  \includegraphics[width=0.7\linewidth]{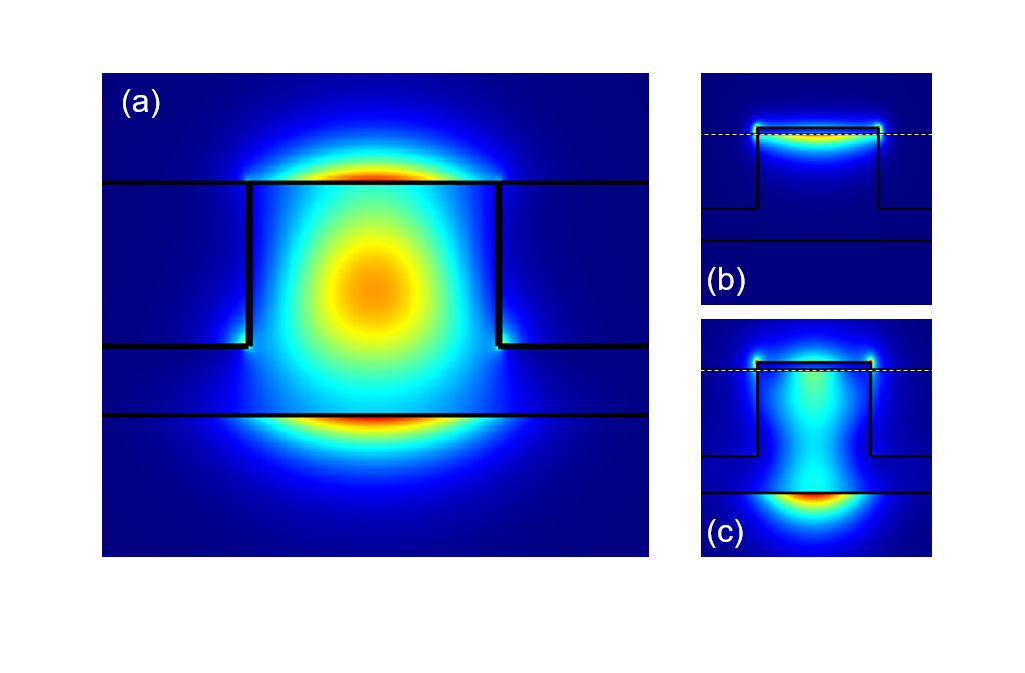}
\caption{(a) Electric field intensity profile of the fundamental TM-mode for a 340$\,$nm$\times$460$\,$nm SOI waveguide. (b) and (c) Electric field intensity profiles of the guided TM-modes for a 340$\,$nm$\times$460$\,$nm SOI waveguide with a graphene sheet and TiN stripe on top of the waveguide at $\lambda=1550\,$nm. The dashed white line represents the graphene sheet plane.}
\label{fig:340example}
\end{figure}

On the other hand, the relative $E$ field intensity at the graphene sheet plane is diminished for waveguide thicknesses above 540$\:$nm, following the trend shown in Fig. \ref{fig:optB}, which is explained by the presence of a less confined TM-mode when the thickness of the photodetector waveguide is increased. However, the previous explanation is inconsistent with the results obtained for 340$\,$–$\:$480$\:$nm  thick waveguides, where the relative $E$ field intensity is not maximized for the smaller thicknesses, as shown in Fig. \ref{fig:optB}. We explain this perceived anomaly by giving further details about the optimization procedure. Fig. \ref{fig:340example}a shows the fundamental TM-mode for a 340$\,$nm$\times$460$\,$nm SOI waveguide, which happens to be the only guided TM-mode. After introducing a TiN stripe on top of the SOI waveguide, we find out that there are two guided TM-modes, which are shown in Fig. \ref{fig:340example}b and \ref{fig:340example}c. For brevity, we call the mode shown in Fig. \ref{fig:340example}b mode (b), and the mode shown in \ref{fig:340example}c mode (c). The coupling efficiency is 23.0\% and 39.9\% for mode (b) and mode (c), respectively. Obviously, the relative $E$ field intensity of mode (b) is greater than that of mode (c), as can be visually discerned. Despite that, in our optimization procedure, we choose mode (c) instead of mode (b), since the coupling efficiency of mode (c) is $\sim 2 \times$ greater than that of mode (b), which means that the power transferred from the fundamental TM-mode of the SOI waveguide to mode (c) is $\sim 2 \times$ larger than it is the case for mode (b). We consistently chose the guided TM-mode in the plasmonic waveguide that yields the highest coupling efficiency with the fundamental TM-mode of the SOI waveguide at each height and width, as previously mentioned. This explains why the relative $E$ field intensity is not maximized for the smaller thicknesses in Fig. \ref{fig:optB}, since the mode with the higher coupling efficiency does not necessarily yield the highest relative $E$ field intensity.

It is also noted in Fig. \ref{fig:optB} that increasing the waveguide width increases the relative $E$ field intensity. The TiN stripe enhances the EM field interaction with graphene by concentrating the EM field at the TiN-graphene interface, and consequently, a wider waveguide with a wider TiN stripe enlarges the interaction area between the propagating mode and the TiN-graphene interface, giving high relative $E$ field intensities. However, inevitable ohmic losses are associated with the strong plasmonic enhancement induced by TiN. Reducing the width of the TiN stripe is a possible workaround to mitigate the ohmic losses, yet the plasmonic enhancement of the EM field at the TiN-graphene interface is diminished for TiN stripes of smaller widths, and a longer photodetector would be required to maintain an overall EM field enhancement that is equivalent to the case of a wide TiN stripe. The effect of the metal stripe width on the device performance was previously investigated for a plasmonic graphene photodetector with TE-polarization \cite{hybrid}, where it has been reported that a wide metal stripe increases ohmic losses, while enhancing the optical absorption of graphene. In this work, we choose the largest stripe width, which is the same as the waveguide width, to achieve the strongest EM field interaction with graphene at the smallest possible footprint, and this comes at the expense of high ohmic losses. For maximum reduction of ohmic losses, a different optimization strategy may be followed, where the stripe width and photodetector length are swept, and the propagation loss is recorded for each of both parameters. The parameters giving the smallest propagation loss, within a given set of design constraints, are the optimum device parameters in such a scenario.

\begin{figure}
\begin{subfigure}{.5\textwidth}
  \centering
  \includegraphics[width=1\linewidth]{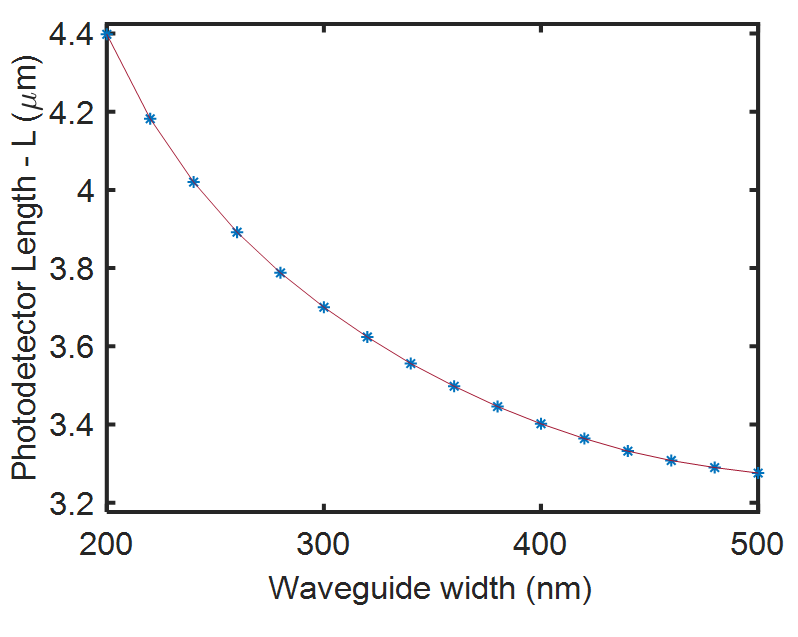}
  \caption{}
  \label{fig:width}
\end{subfigure}%
\begin{subfigure}{.5\textwidth}
  \centering
  \includegraphics[width=1\linewidth]{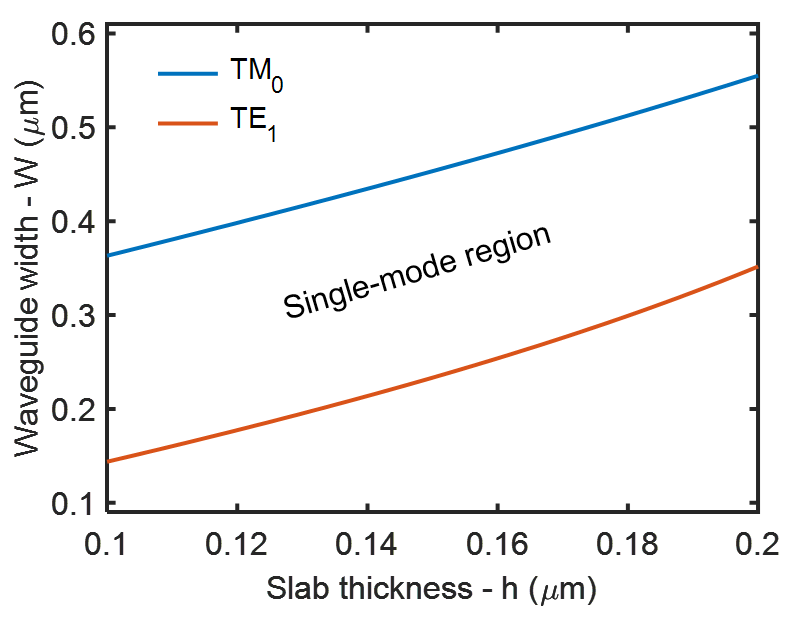}
  \caption{}
  \label{fig:cutoff}
\end{subfigure}
\caption{(a) Photodetector length and the corresponding waveguide width for a fixed absorption of 95\%. (b) Cutoff curves for single-mode operation in a deep-etched silicon rib waveguide with a 460$\:$nm thick waveguide at $\lambda=1550\,$nm.}
\label{fig:correction}
\end{figure}

The absorbed optical power depends on the device geometry. Fig. \ref{fig:width} shows the lengths and widths at which 95\% of the power is absorbed. One can shorten the photodetector length and maintain the same power absorption when the waveguide width is simultaneously increased. However, for widths beyond 460$\,$nm, the length reduces at a lower rate than the width, which would make further reduction in length meritless in terms of the overall device footprint. Based on that, the optimum device geometry is $460\times460\,$nm, where 95\% of the power is absorbed at 3.3$\,$\textmu m length. However, a $460\times460\,$nm geometry with a 100$\:$nm thick slab does not satisfy the single-mode condition for deep-etched sub-micron SOI rib waveguides at $\lambda = 1550\:$nm, which is determined by the following relations \cite{sc}:

\begin{equation} \label{condition1}
    W/H < \dfrac{(55.679H^2 -45.467H + 10.737)r}{\sqrt{1-r^2}} + (21.805H^2 - 21.473H + 5.696)
\end{equation} 

\begin{equation} \label{condition2}
    W/H > \dfrac{(-14.905H^2 + 8.638H + 0.581)r}{\sqrt{1-2r^2}} + (22.283H^2 - 19.310H + 4.160)
\end{equation}

\begin{equation}\label{eq:etch}
    r = h/H\; , \; \; r < 0.5
\end{equation}

Where $W$ is the waveguide width, $H$ is the waveguide height, and $h$ is the slab thickness. A silicon rib waveguide is considered deep-etched when $h<H/2$, which is expressed as $r<0.5$ in Eq. \ref{eq:etch}, and the herein proposed waveguide structure is deep-etched based on this definition. The cutoff curves for the single-mode condition are plotted in Fig. \ref{fig:cutoff}, for a fixed 460$\:$nm height. The single-mode condition is satisfied by replacing the 100$\:$nm thick slab with a 160$\:$nm slab for 460$\:$nm and 500$\:$nm thick waveguides. The single-mode condition is also satisfied for an etch depth variation of $\pm$5$\:$nm, which is a routinely obtained precision for 300$\:$mm wafers. For such wafers, statistical variations ($3\sigma$) as low as $\pm$7.6$\:$nm linewidth uniformity and $\pm$1$\:$nm thickness uniformity can be achieved using state-of-the-art fabrication technologies employed in silicon photonics \cite{sweetSpot}. Variations of that order have a negligible impact on the photodetector operation based on the data that we computed for the coupling efficiency and the relative $E$ field intensity, while a width variation of $\pm$10$\:$nm results in a device length variation of $<$10$\:$nm considering a 460$\:$nm target width; the resulting variation is $\sim$2$\times$ orders of magnitude less than the photodetector length which is on the order of a few microns. Thanks to the mature CMOS-based fabrication technologies and the phase-insensitive operation of the photodetector, typical variations in the SOI waveguide geometry are not of substantial concern for this device. Interestingly, our study reveals that the device performance is more sensitive to the graphene sample quality and its preparation method (see section 4 of the main text). 

\begin{figure}
  \centering
  \includegraphics[width=0.5\linewidth]{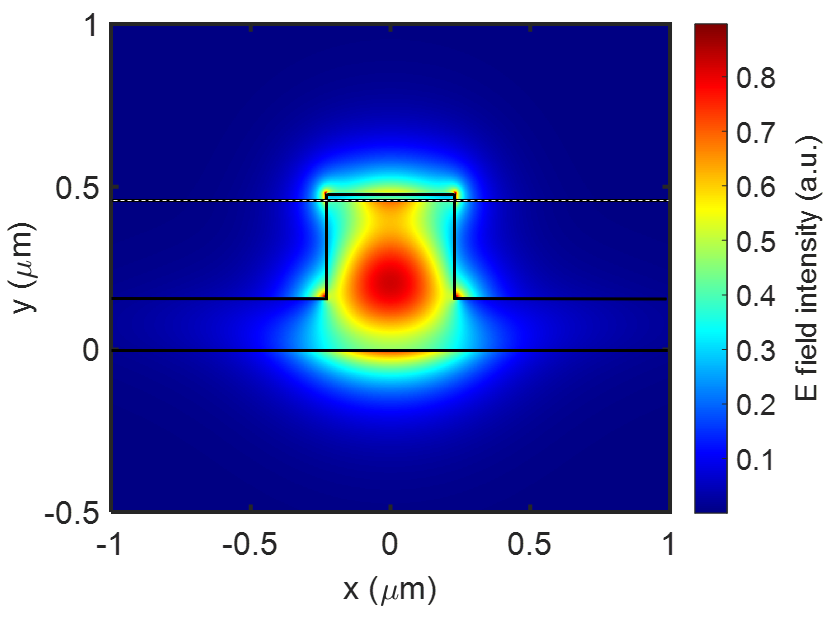}
  \caption{Electric field intensity profile of the propagating TM-mode in the waveguide-integrated photodetector at $\lambda=1550\,$nm. The dashed white line represents the graphene sheet plane.}
  \label{fig:mode}
\end{figure}

Similar coupling efficiencies were computed after increasing the slab thickness to 160$\:$nm, namely 79\% and 84\% for the 460$\:$nm and 500$\:$nm thick waveguides, respectively. The computed TM-mode for the optimum waveguide geometry is shown in Fig. \ref{fig:mode}. By taking into account that the TiN film may be misaligned by a typical $\sim$$20\:$nm stitching error \cite{956140, doi:10.1116/1.3700439}, the computed coupling efficiencies degrade to 77.2\% and 82.6\% for the 460$\:$nm and 500$\:$nm thick waveguides, respectively. Thin TiN films can be made with thicknesses as small as a few nanometers by atomic layer deposition \cite{VANBUI201345}, which makes it possible to deposit 20$\:$nm thick TiN films with relatively high precision. With a 160$\:$nm thick slab, 95\% of the power is absorbed at 3.5$\:$\textmu m length for the optimum $460\times460\:$nm waveguide geometry.

\begin{figure}[H]
  \centering
  \includegraphics[width=0.8\linewidth]{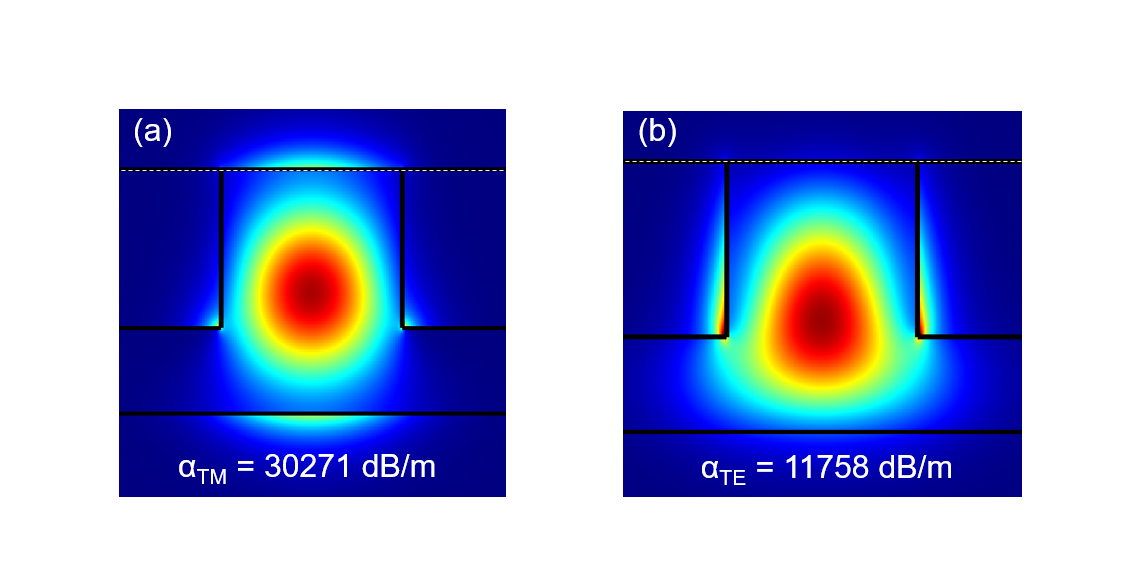}
  \caption{Fundamental (a) TM-mode and (b) TE-mode of a 460$\,$nm$\times$460$\,$nm SOI waveguide with a graphene sheet placed on top of the waveguide. At $\lambda=1550\,$nm, the optical absorption of graphene is 30271$\,$dB/m and 11758$\,$dB/m for the TM-mode and TE-mode, respectively}
  \label{TMte}
  \end{figure}

The TM-mode was chosen for its superior interaction with the graphene sheet, which yields a $\sim$$2.5\times$ higher absorption efficiency than the transverse electric (TE)-mode, as shown in Fig. \ref{TMte}. This strong absorption is due to the impinging $E$ field on the graphene sheet that is placed on top of the waveguide for the TM-mode, which is unlike the case for the TE-mode where the $E$ field is only parallel to the graphene sheet, and interacts less intensely with graphene as a result. This finding is corroborated by several theoretical and experimental studies as reported in \cite{study1, all-optical, opticsComm}. Moreover, the propagating TM-mode experiences a plasmonic near-field enhancement when the TiN stripe is placed on top of the graphene sheet, which further enhances the effective absorption of graphene, since the TM-mode can induce a stronger plasmonic response in TiN than a TE-mode; this is also the case for any metal \cite{maier2007plasmonics}. This is in contrast with plasmonic slot waveguides reported in \cite{plasmonic3, plasmonic4}, where the TE-mode does not experience a similarly strong plasmonic near-field enhancement, since graphene interacts less effectively with TE-modes than it would with TM-modes. Similar arguments can be made for the bowtie nanostructure reported in \cite{plasmonic1}. While in \cite{plasmonic2}, the graphene sheet and the metal are both placed on top of the waveguide, yet the reported device supports a TE-mode, which limits the induced plasmonic near-field enhancement. It can be seen in Fig. \ref{TMte} (a) that the absorption of the TM-mode is 0.0303$\,$dB/$\text{\textmu}$m, which corresponds to an absorption of $\sim2.4$\% for a 3.5$\text{\textmu}$m long device. Such is the absorption of the graphene-silicon waveguide without the TiN stripe, which is coincidentally similar to the intrinsic absorption of graphene in the case of vertically incident light ($\sim$$2.3$\%). As was explained in the Methods section of the main text, the plasmon-enhanced effective absorption of graphene with the TiN stripe was $A_G = 5.2$\%. Therefore, we observe a $>$$2\times$ enhancement of graphene's absorption, purely because of the presence of the TiN stripe.

A thick TiN stripe introduces a strong plasmonic enhancement of the propagating TM-mode at the graphene sheet plane, yet the coupling efficiency deteriorates for a thick stripe, e.g. 61.5\% for a 30$\,$nm thick stripe. On the other hand, the use of a thin TiN stripe leads to a high coupling efficiency, e.g. 94\% for a 10$\,$nm thick stripe, yet unsatisfactory plasmonic enhancement may be associated with thin plasmonic films, since the plasmonic enhancement strongly depends on the film thickness \cite{doi:10.1021/acs.jpcc.9b06592}, besides there are challenges involved with the deposition of ultra-thin continuous metal films \cite{doi:10.1021/acsphotonics.9b00907}. Therefore, a 20$\:$nm thick stripe is chosen in this work as a tradeoff between both extremes. 

  \begin{figure}
  \centering
  \includegraphics[width=0.5\linewidth]{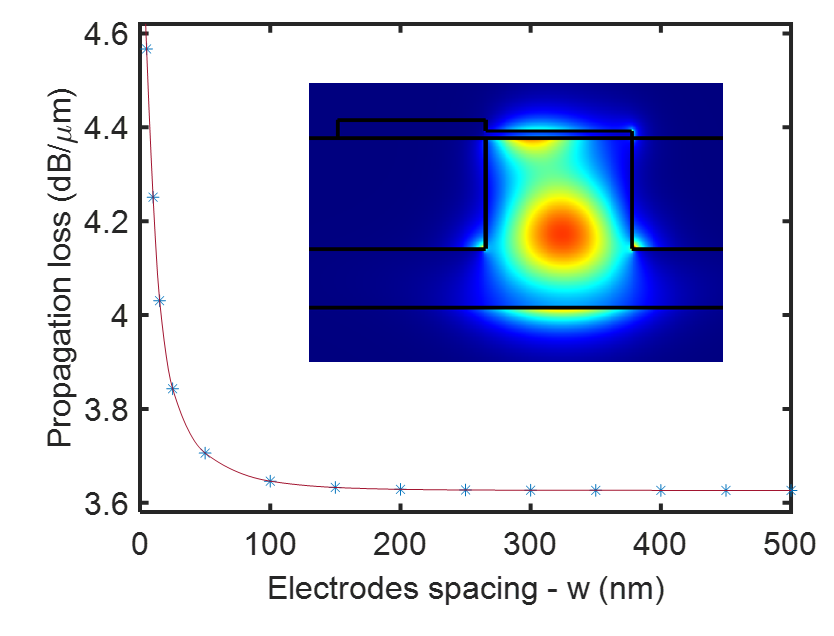}
  \caption{Propagation loss as a function of the electrodes spacing at $\lambda=1550\,$nm. The side electrode is 50$\,$nm thick. Inset illustrates the effect of placing the electrodes near each other.}
  \label{fig:spacing}
  \end{figure}

It is desirable to have the two TiN electrodes placed near each in order to maximize the intensity profile, $I(x)$, across the graphene sheet. The carrier heat is proportional to $I(x)$, which decreases as the electrodes spacing increases (see section 4). However, placing the side TiN electrode in close proximity to the waveguide deforms the propagating mode and introduces ohmic losses, most notably for distances below 50$\,$nm (see Fig. \ref{fig:spacing}), in addition to a reduced net difference in the carrier temperature (and consequently $V_{PTE}$) as the side electrode approaches the device center (see Fig. 5 of the main text). To avoid such issues, we place the side electrode 200$\,$nm away from the waveguide edge, where the loss curve exhibits a plateau, and the impact of the side electrode on the propagating mode becomes negligible, in addition to achieving a high intensity profile and a high net difference in the carrier temperature, resulting in a high net PTE voltage.

\section{Intraband and Interband Transitions}

\begin{figure}
\begin{subfigure}{.5\textwidth}
  \centering
  \includegraphics[width=0.8\linewidth]{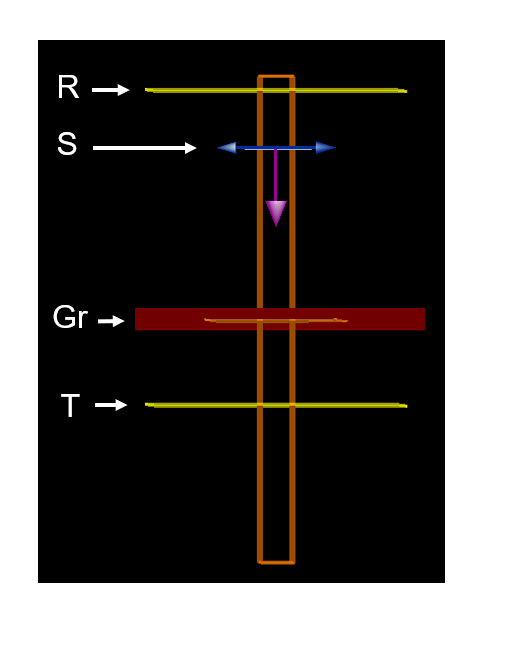}
  \caption{}
  \label{fig:setup}
\end{subfigure}%
\begin{subfigure}{.5\textwidth}
  \centering
  \includegraphics[width=1\linewidth]{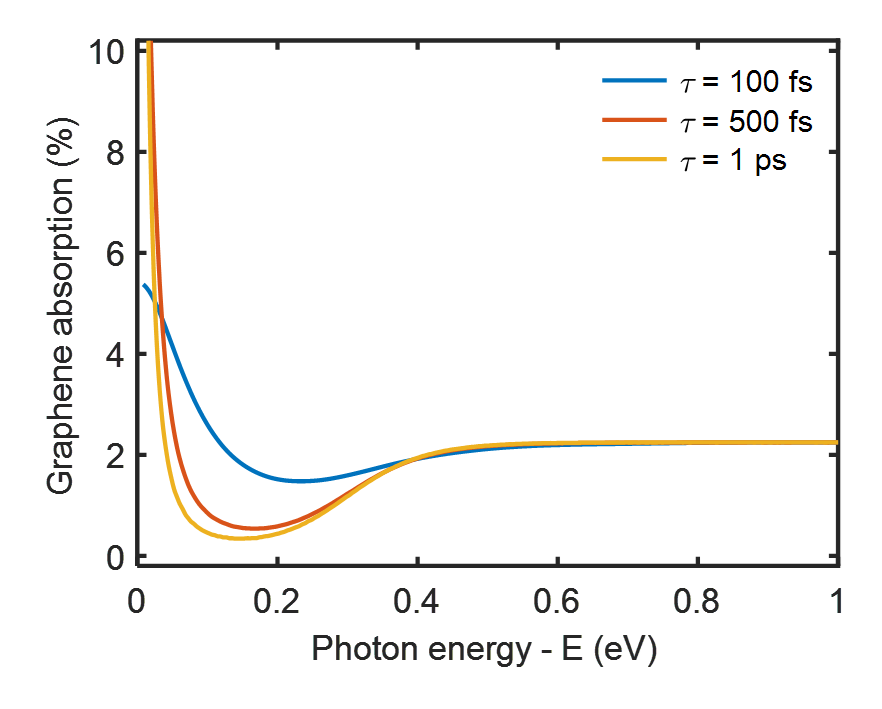}
  \caption{}
  \label{fig:absorption}
\end{subfigure}
\caption{(a) Simulation setup. R is a reflection monitor, S is a broadband plane wave source, Gr is graphene, and T is a transmission monitor. (b) Graphene absorption as a function of the incident photon energy for multiple scattering rates.}
\end{figure}

To find out how the scattering rate and incident wavelength of light affect the absorption of graphene, we run a simple FDTD simulation of the schematic shown in Fig. \ref{fig:setup}. In the FDTD simulation, a broadband plane wave source is incident on the graphene sheet, and the optical absorption of graphene (A) is extracted from the reflection (R) and transmission (T) coefficients:

\begin{equation}
    A = 1 - T - R
\end{equation}

The chemical potential of graphene is fixed at $\text{\textmu}=0.15\,$eV. The resultant absorption spectrum of graphene is shown in Fig. \ref{fig:absorption}. We notice that scattering time ($\tau$) variations have a significant impact on the optical absorption of graphene for photon energies $<0.3\,$eV. However, starting from $\text{\textmu}=0.3\,$eV, the optical absorption of graphene is less sensitive to variations in $\tau$. We find out that the optical absorption of graphene is practically insensitive to changes in $\tau$ at $\lambda=1550\,$nm ($E=0.8\,$eV), where the absorption is simply $\sim2.3$\%. Hence, we conclude that for a graphene sheet with $\text{\textmu}=0.15\,$eV, the optical absorption is determined by interband transitions for incident photons with an energy $E=0.8\,$eV. This conclusion is consistent with the optoelectronic properties of graphene described in \cite{grapheneBook}, where interband transitions start dominating the absorption of graphene when $\hbar\omega > 2\text{\textmu}$.

\section{Graphene Absorption vs Wavelength}

\label{secWavelength}
\begin{figure}
  \centering
  \includegraphics[width=0.5\linewidth]{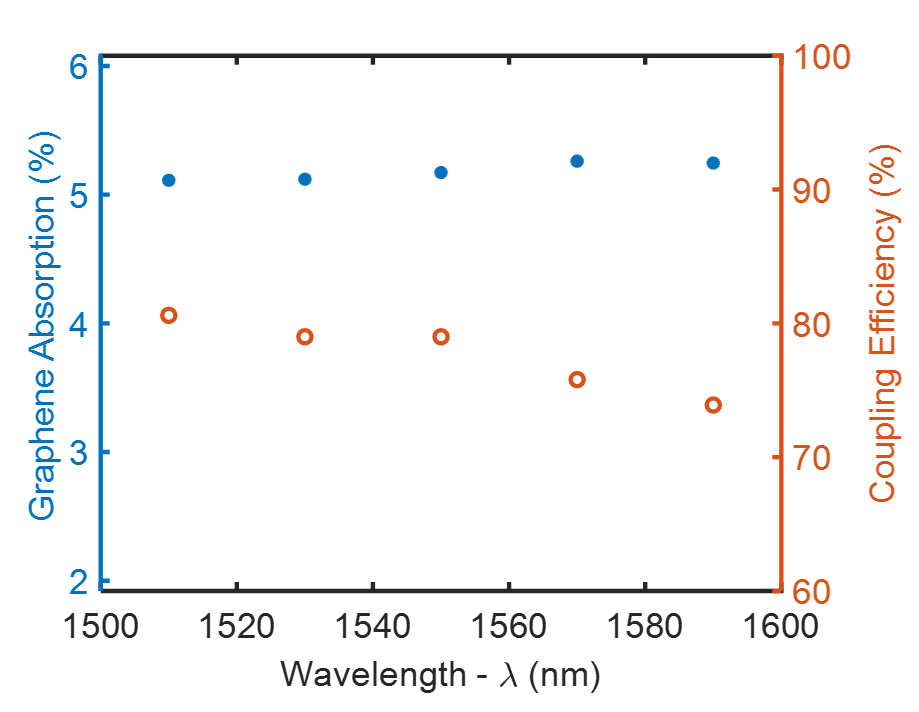}
  \caption{Effective absorption of graphene and coupling efficiency as a function of wavelength.}
  \label{fig:grapheneWavelength}
  \end{figure}

Following the model described in the Methods section of the main text, we computed the effective absorption of graphene at other wavelengths in the telecom C-band. It is concluded that the effective absorption slightly increases at longer wavelengths (see Fig. \ref{fig:grapheneWavelength}), since the propagating mode becomes more confined in the photodetector waveguide, and the graphene sheet absorbs more of the propagating mode as a result. The coupling efficiency is also shown in Fig. \ref{fig:grapheneWavelength}. It is clearly noted that the coupling efficiency decreases at a higher rate than the effective absorption at longer wavelengths in this band, and vice versa.

Table S1 presents the refractive indices of TiN used in this study. The refractive index values are based on the dielectric constant values reported in \cite{gosciniak_justice_khan_corbett_2016, embedded}, where thin 30 nm-thick films of TiN were deposited on n-Si/p-Si substrates by DC reactive magnetron sputtering from a 99.99\% titanium target in an Argon-Nitrogen environment. The deposition rate and substrate temperature were kept constant at 1.38 nm/min and 150\textdegree C, respectively.

\begin{table}
  \caption{Refractive index of titanium nitride}
  \label{tbl:example}
  \begin{tabular}{ll}
    \hline
    $\lambda$ (nm)  & $\:\:\:\:\:\:\:$n+jk  \\
    \hline
    1510 & 2.50 + j7.58 \\
    1530 & 2.52 + j7.72 \\
    1550 & 2.54 + j7.84 \\
    1570 & 2.58 + j7.94 \\
    1590 & 2.65 + j8.04 \\
    \hline
  \end{tabular}
\end{table}

\section{Intensity Profile}

Light intensity, or the irradiance ($I$), is defined as the power received by a surface per unit area, and has the units (W/m$^2$). In our case, the surface of interest is the graphene sheet surface, located on top of the silicon waveguide and beneath the TiN stripe. We can figure out the portion of optical power absorbed by graphene in the $z$ direction, or the direction of propagation by using the Lumerical FDTD power monitors, as was presented in the Methods section of the main text. Then by taking the power difference across the photodetector length, we end up with a one-dimensional (1D) power profile along the photodetector width, which can be used to solve the 1D heat transport equation given in \cite{acshotcarrier, High-Responsivity, asymmetric}. As was explained in the main text, the carrier temperature profile is calculated using the analytical solution to the heat transport equation \cite{cooling, jacek}:

\begin{equation} \label{eq:analytical}
    \Delta T = T_{c}(x) - T_{0} = \dfrac{\xi sinh ((x_{0} - |x|)/ \xi)}{2 cosh(x_{0}/\xi)} \left( \dfrac{A_{G} I(x)}{\kappa} \right) 
\end{equation}

The intensity profile plugged into the analytical solution to the heat transport equation is a 1D power profile per unit length:

\begin{equation}
    I(x) = \dfrac{\mathcal{P}(x)}{x_{0}}
\end{equation}

Where $\mathcal{P}(x)$ is the power profile of the excitation waveguide mode. Plugging $I(x)$ into Eq. \ref{eq:analytical} gives a temperature profile ($\Delta T$) in units of Kelvin, since $I(x)$ has the units of (W/m). Finding out the power profile of the computed waveguide modes in Lumerical can be tricky, since the powers of the injected waveguide modes are arbitrary. We know that almost all of the power injected into the photodetector waveguide is absorbed across the photodetector length, or 95\% of the power to be exact. Therefore, it is sufficient to multiply the input power profile by 95\% in order to find out the power absorbed across the photodetector length. We extract the real part of the Poynting vector profile ($P$) from a power monitor placed 10$\,$nm away from the mode source in Lumerical FDTD; the power fraction transmitted through the power monitor is almost unity (99\%). We then use the extracted Poynting vector profile as a substitute for the power profile. Both the power and the Poynting vector share a similar profile, as the time-averaged power flowing across a surface is related to the Poynting vector by the following relation \cite{poynting}: 

\begin{equation}
    \mathcal{P} = \dfrac{1}{2} \int_{S} Re(P) \: dS = \dfrac{1}{2} * Area * Re(P)
\end{equation}

\begin{figure}
\begin{subfigure}{.5\textwidth}
  \centering
  \includegraphics[width=1\linewidth]{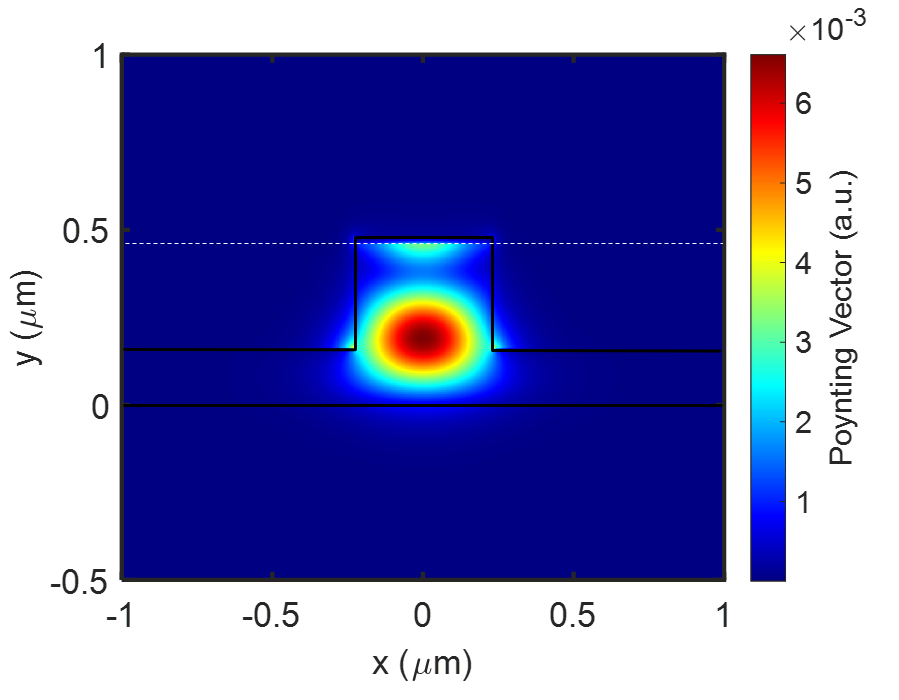}
  \caption{}
  \label{fig:Poynting1}
\end{subfigure}%
\begin{subfigure}{.5\textwidth}
  \centering
  \includegraphics[width=1\linewidth]{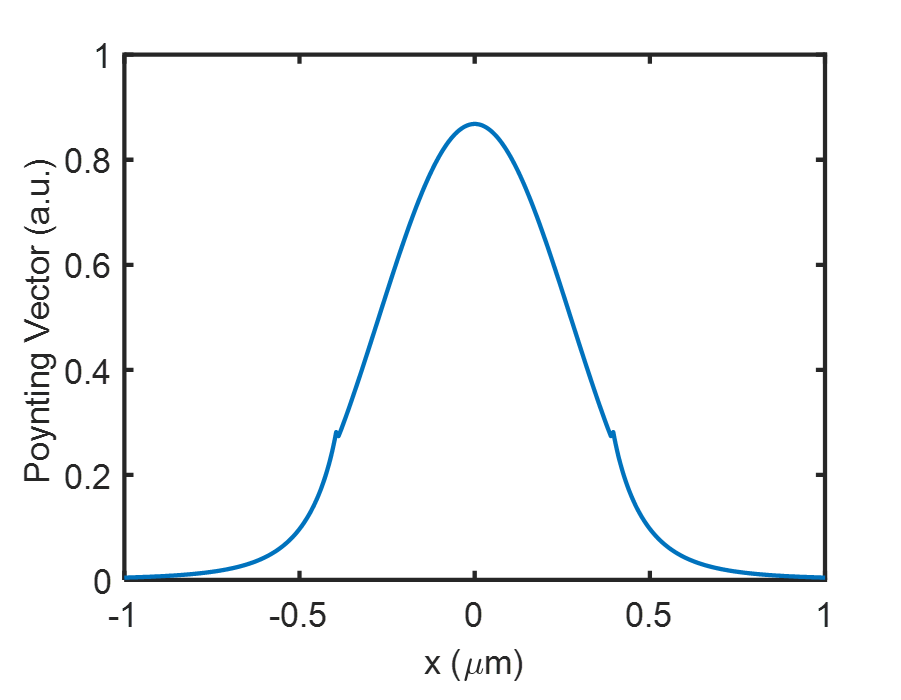}
  \caption{}
  \label{fig:Poynting2}
\end{subfigure}
\caption{(a) 2D and (b) 1D profile of the Poynting vector in the photodetector waveguide at $\lambda = 1550\,$nm. The white dashed line represents the graphene sheet plane.}
\end{figure}

Where $S$ denotes a surface integral. In other words, the power is equal to the real part of the Poynting vector times a scaling factor: (1/2)*Area. Replacing $\mathcal{P}$ with $P$ is further justified since the values of both the Poynting vector and the power of the imported mode are arbitrary. Nonetheless, the profile shapes of both $\mathcal{P}$ and $P$ are physical. Fig \ref{fig:Poynting1} shows the distribution profile of the Poynting vector in the photodetector waveguide.

The 1D Poynting vector, $P(x)$, can be computed by integrating the distribution profile of the Poynting vector across all values of $y$:

\begin{equation}
    P(x) = \int_{-\infty}^{+\infty} P(x,y) \: dy
\end{equation}

Fig \ref{fig:Poynting2} shows the resultant $P(x)$ for the Poynting vector profile of Fig \ref{fig:Poynting1}. It is noted in Fig \ref{fig:Poynting2} that the maximum value of $P(x)$ is located at the waveguide center. Therefore, $x_{0}$ is taken as the distance from the center of the silicon waveguide to the external TiN electrode, namely 660$\,$nm. Since we are replacing $\mathcal{P}(x)$ with $P(x)$, $I(x)$ can now be calculated by dividing $P(x)$ by $x_{0}$, after dividing $P(x)$ by 2, because the external electrode is placed on one side of the photodetector. Unlike photovoltaic photodetectors, where placing electrodes on both sides of the photodetector induces flow of both carrier types, namely electrons and holes, across the photodetector, thus doubling the photocurrent; placing two side electrodes for a graphene PTE photodetector does not affect the responsivity, such that when two side electrodes are present, the resultant PTE voltage induces a photocurrent on each side of the photodetector with half the original photocurrent value, based on elementary circuit theory.

Fortunately, we are interested in the responsivity of the photodetector, not the photo-induced PTE voltage. Therefore, we divide the resultant arbitrary PTE voltage by the arbitrary total input power in the waveguide ($P_{total}$):

\begin{equation}
    P_{total} = \int_{-\infty}^{+\infty}\int_{-\infty}^{+\infty} P(x,y) \: dxdy
\end{equation}

The resultant responsivity is physical, since the PTE voltage was calculated based on the same input power profile. This method was repeated for each of the wavelengths presented in this study.

\bibliography{achemso-demo}

\end{singlespace}